\documentclass{report}
\usepackage{array}
\usepackage{citesort}
\usepackage{amssymb,amsmath}
\usepackage{multirow}
\usepackage{epsfig}
\usepackage[round]{natbib}

\newcommand {\mm}[1] {\ifmmode{#1}\else{\mbox{\(#1\)}}\fi}
\newcommand{\real} {\mm{{\Bbb R}}}

\newcommand{\utwi}[1]{\mbox{\boldmath $ #1$}}
\newcommand{\ba}{{\utwi{a}}}

\newcommand{\bc}{{\utwi{c}}}

\newcommand{\bs}{{\utwi{s}}}

\newcommand{\bu}{{\utwi{u}}}
\newcommand{\bv}{{\utwi{v}}}
\newcommand{\bw}{{\utwi{w}}}
\newcommand{\bx}{{\utwi{x}}}
\newcommand{\by}{{\utwi{y}}}
\newcommand{\bz}{{\utwi{z}}}

\newcommand{\bM}{{\utwi{M}}}

\newcommand{\bS}{{\utwi{S}}}

\def\i{\mbox{\boldmath $i$}}
\def\j{\mbox{\boldmath $j$}}

\title{Chapter 4: Knowledge-based energy functions for computational
studies of proteins} 
\author{ Xiang Li and Jie Liang\\ Dept of Bioengineering, University of Illinois at Chicago, Chicago, IL, 60607\\To be published in a book by Springer}
\begin{document}

\maketitle
\tableofcontents

\section{Introduction}
This chapter discusses theoretical framework and methods for
developing knowledge-based potential functions essential for protein
structure prediction, protein-protein interaction, and protein
sequence design.  We discuss in some details about the
Miyazawa-Jernigan contact statistical potential,
distance-dependent statistical potentials, as well as geometric
statistical potentials.  We also describe a geometric model for
developing both linear and non-linear potential functions by
optimization. Applications of knowledge-based potential functions in
protein-decoy discrimination, in protein-protein interactions, and
in protein design are then described. Several issues of knowledge-based
potential functions are finally discussed.

In the experimental work that led to the recognition of the 1972
Nobel prize in chemistry, Christian Anfinsen showed that a
completely unfolded protein ribonuclease could refold
spontaneously to its biologically active conformation. This
observation indicates that the sequence of amino acids of a
protein contains all of the information needed to specify its
three-dimensional
structure~\citep{Anfinsen61_PNAS,Anfinsen73_Science}. The
automatic {\it in vitro\/} refolding of denatured proteins was
further confirmed in many other protein
systems~\citep{ThermoHypothesisReview}.  Anfinsen's experiments
led to the thermodynamic hypothesis of protein folding, which
postulates that a native protein folds into a three-dimensional
structure in equilibrium, in which the state of the whole
protein-solvent system corresponds to the global minimum of free
energy under physiological conditions.

Based on this thermodynamic hypothesis, computational studies of
proteins, including structure prediction, folding simulation, and
protein design, all depend on the use of a potential function for
calculating the effective energy of the molecule. In protein structure
prediction, the potential function is used either to guide the
conformational search process, or to select a structure from a set of
possible sampled candidate structures. Potential function has been
developed through an inductive approach~\citep{Sippl93_JCAMD}, where
the parameters are derived by matching the results from
quantum-mechanical calculations on small molecules to experimentally
measured thermodynamic properties of simple molecular systems. These
potential functions are then generalized to the macromolecular level
based on the assumption that the complex phenomena of macromolecular
systems result from the combination of a large number of interactions
as found in the most basic molecular systems. This type of potential
function is often referred to as ``physics-based'', ``semi-empirical''
effective potential function, or a force
field~\citep{Levitt75_Nature,Wolynes95_Science,Scheraga75_JPC,Karplus90_Nature}.
The physics-based potential functions have been extensively studied
over the last three decades, and has found wide uses in protein
folding
studies~\citep{Duan&Kollman98_Science,Lazaridis&Karplus00_COSB}.
Nevertheless, it is difficult to use physics-based potential functions
for protein structure prediction, because they are based on full
atomic model and therefore require high computational cost. In
addition, a physical model may not fully capture all of the important
physical interactions.  Readers are referred to Chapter 4 for a
detailed discussion of physics-based potential functions.

Another type of potential function is developed through a deductive
approach by extracting the parameters of the potential functions from
a database of known protein structures~\citep{Sippl93_JCAMD}. Because
this approach implicitly incorporates many physical interactions
(electrostatic, van der Walls, cation-$\pi$ interactions) and the
extracted potentials do not necessarily reflect true energies, it is
often referred to as ``knowledge-based effective energy function''. In
recent past, this approach has quickly gained momentum due to the
rapidly growing database of experimentally determined
three-dimensional protein structures.  Impressive successes in protein
folding, protein-protein docking and protein design have been achieved
recently using knowledge-based scoring
functions~\citep{proteinDesignReview,CASP,Hu&Liang04_Bioinformatics,CAPRI}.
In this chapter, we focus our discussion on this type of potential
functions.

\section{General framework}
Several different approaches have been proposed to extract
knowledge-based scoring functions from protein structures. They can be
categorized roughly into two groups. One prominent group of
knowledge-based potentials are those derived from statistical analysis
of database of protein structures
\citep{TanakarScheraga76,Miyazawa&Jernigan85_M,Samudrala&Moult98_JMB,Lu&Skolnick01_Proteins}.
In this class of potentials, the interacting potential between a
pair of residues are estimated from its relative frequency in
database when compared with that in a reference state or a null
model
\citep{Miyazawa&Jernigan96_JMB,Samudrala&Moult98_JMB,Lu&Skolnick01_Proteins,Wodak93_COSB,Sippl95_COSB,Lerner95_Proteins,Jernigan96_COSB,Simons99_Proteins}.
A different class of knowledge-based potentials are based on
optimization. In this case, the set of parameters for the
potential functions are optimized by some criterion, {\it e.g.},
by maximizing the energy gap between known native conformation and
a set of alternative (or decoy) conformations
\citep{Goldstein92_PNAS,Maiorov&Crippen92_JMB,Thomas&Dill96_PNAS,Tobi&Elber00_Proteins_1,Vendruscolo98_JCP,Vendruscolo00_Proteins,Bastolla01_Proteins,Dima00_PS,Micheletti01_Proteins,Dobbs02_Proteins,Hu&Liang04_Bioinformatics}.

There are three main ingredients for developing a
knowledge-based potential function. We first need {\it protein
descriptors\/} to describe the sequence and the shape of the
native protein structure in a format that is suitable for
computation. We then need to decide on a {\it functional form\/}
of the potential function. Finally, we need a {\it method to
derive the values of the parameters\/} for the potential function.

\subsection{Protein representation and descriptors.}
To describe the geometric shape of a protein and its sequence of
amino acid residues, a protein is frequently represented by a
$d$-dimensional descriptor $\bc \in \real^d$. For example, a
method that is widely used is to count non-bonded contacts of 210
types of amino acid residue pairs in a protein structure. In this
case, the count vector $\bc \in \real^d, d=210$, is used as the
protein descriptor. Once the structural conformation of a protein
$\bs$ and its amino acid sequence $\ba$ is given, the protein
descriptions $f: (\bs, \ba) \mapsto \real^d$ will fully determine
the $d$-dimensional vector $\bc$. In the case of contact
descriptor, $f$ corresponds to the mapping provided by specific
contact definition, {\it e.g.}, two residues are in contact if
their distance is below a cut-off threshold distance. At the
residue level, the coordinates of of $C_{\alpha}$, $C_{\beta}$, or
side-chain center can be used to represent the location of a
residue. At atom level, the coordinates of atoms are directly
used, and contact may be defined by the spatial proximity of
atoms.  In addition, other features of protein structures can be
used as protein descriptors as well, including distances between
residue or atom pairs, solvent accessible surface areas, dihedral
angles of backbones and side-chains, and packing densities.

\subsection{Functional form.}
The form of the potential function $H:\real^d \mapsto \real$
determines the mapping of a $d$-dimensional descriptor $\bc$ to a
real energy value. A widely used functional form for protein
scoring function $H$ is the weighted linear sum of pairwise
contacts
\citep{TanakarScheraga76,Miyazawa&Jernigan85_M,Tobi&Elber00_Proteins_1,Vendruscolo98_JCP,Samudrala&Moult98_JMB,Lu&Skolnick01_Proteins}.
The linear sum $H$ is:
\begin{equation}
H(f(\bs, \ba)) = H(\bc) = \bw \cdot \bc = \sum_i w_i c_i,
\label{linear}
\end{equation}
where ``$\cdot$'' denotes inner product of vectors; $c_i$ is the
number of occurrence of the $i$-th type of descriptor. As soon as
the weight vector $\bw$ is specified, the potential function is
fully defined. In subsection~\ref{subsec:nonlinear},
we will discuss a nonlinear form potential function.

\subsection{Deriving parameters of potential functions.}
For statistical knowledge-based potential functions, the weight vector
$\bw$ for linear potential is derived by characterization of the
frequency distributions of structural descriptors from a database
of experimentally determined protein structures. For optimized
knowledge-based linear potential function, $\bw$ is obtained through
optimization. We describe the details of these two approaches
below.

\section{Statistical method}

\subsection{Background}
In statistical methods, the observed statistical frequencies of
various protein structural features are converted into effective
free energies, based on the assumption that frequently observed
structural features correspond to low-energy
states~\citep{Scheraga76_M,Miyazawa&Jernigan85_M,Sippl90_JMB}. This
is the Boltzmann's principle, an idea first proposed by Tanaka and
Scheraga (1976) to estimate potentials for pairwise interaction
between amino acids~\citep{Scheraga76_M}. Miyazawa and Jernigan
(1985) significantly extended this idea and derived a widely-used
statistical potentials, where solvent terms are explicitly
considered and the interactions between amino acids are modeled by
contact potentials. Sippl (1990) and
others~\citep{Samudrala&Moult98_JMB,Lu&Skolnick01_Proteins,Zhou02_ProSci}
derived distance-dependent energy functions to incorporate both
short-range and long-range pairwise interactions. The pairwise
terms were further augmented by incorporating dihedral
angles~\citep{Nishikawa&Matsuo93_ProEng,Kocher&Wodak94_JMB},
solvent accessibility and
hydrogen-bonding~\citep{Nishikawa&Matsuo93_ProEng}. Singh and
Tropsha (1996) derived potentials for higher-order
interactions~\citep{Singh&Tropsha96_JCB,Singh96_JCB}. More
recently, Ben-Naim (1997) presented three theoretical examples to
demonstrate the nonadditivity of three-body
interactions~\citep{Ben-Naim97_JCP}. Li and Liang (2005) identified
three-body interactions in native proteins based on an accurate
geometric model, and quantified systematically the nonadditivities
of three-body interactions~\citep{Li&Liang05_Proteins}.

\subsection{Theoretical model.}  At the equilibrium state, an
individual molecule may adopt many different conformations or
microscopic states with different probabilities. The distribution
of protein molecules among the microscopic states follows the
Boltzmann distribution, which connects the potential function
$H(\bc)$ for a microstate $\bc$ to its {\it probability of
occupancy $\pi(\bc)$ }. This probability $\pi(\bc)$ or the
Boltzmann factor is:
\begin{equation}
\pi(\bc) = \exp[-H(\bc)/kT]/Z(\ba), \label{eq:20}
\end{equation}
where $k$ and $T$ are Boltzmann constant and the absolute
temperature measured in Kelvin, respectively. The partition
function $Z(\ba)$ is defined as:
\begin{equation}
Z(\ba) \equiv \sum_{\bc} \exp[-H(\bc)/kT].
\end{equation}
It is a constant under the true energy function once the sequence
$\ba$ of a protein is specified, and is independent of the
representation $f(\bs, \ba)$ and descriptor $\bc$ of the protein.
If we are able to measure the probability distribution $\pi(\bc)$
accurately, we can obtain the knowledge-based potential function
$H(\bc)$ from the Boltzmann distribution:
\begin{equation}
H(\bc) = -kT \ln \pi(\bc) - kT \ln Z(\ba). \label{eq:21}
\end{equation}
The partition function $Z(\ba)$ cannot be obtained directly from
experimental measurements. However, at a fixed temperature,
$Z(\ba)$ is a constant and has no effect on the different
probability of occupancy for different conformations.

In order to obtain an knowledge-based potential function that encodes
the sequence-structure relationship of proteins, we have to remove
background energetic interactions $H'(\bc)$ that are independent
of the protein sequence and the protein structure. These generic
energetic contributions are referred collectively as that of the
{\it reference state\/} \citep{Sippl90_JMB}. An {\it effective
potential energy\/} $\Delta H(\bc)$ is then obtained as:
\begin{equation}
\Delta H(\bc) = H(\bc) - H'(\bc)
 =  -kT \ln [\frac{\pi(\bc)}{\pi'(\bc)}] -kT \ln
[\frac{Z(\ba)}{Z'(\ba)}],
\label{eq:Z}
\end{equation}
where $\pi'(\bc)$ is the probability of a sequence adopting a
conformation specified by the vector $\bc$ in the reference state.
Since $Z(\ba)$ and $Z'(\ba)$ are both constants, $-kT \ln
(Z(\ba)/Z'(\ba))$ is also a constant that does not depend on the
descriptor vector $\bc$. If we assume that $Z(\ba) \approx
Z'(\ba)$ as in~\citep{Sippl90_JMB}, the effective potential energy
can be calculated as:
\begin{equation}
\Delta H(\bc) = -kT \ln [\frac{\pi(\bc)}{\pi'(\bc)}]
\label{eq:Sippl0}
\end{equation}

To calculate $\pi(\bc)/\pi'(\bc)$, One can further assume that the
probability distribution of each descriptor is independent, and we
have $\pi(\bc)/\pi'(\bc) = \prod_i [\frac{\pi(c_i)}{\pi'(c_i)}].$
Furthermore, by assuming each occurrence of the $i$-th descriptor
is independent, we have $\prod_i [\frac{\pi(c_i)}{\pi'(c_i)}] =
\prod_i \prod_{c_i} [\frac{\pi_i}{\pi'_i}],$ where $\pi_i$ and
$\pi'_i$ are the probability of $i$-th type structural feature in
native proteins and the reference state, respectively. In a linear
potential function, the right-hand side of
Equation~\ref{eq:Sippl0} can be calculated as:
\begin{equation}
-kT \ln [\frac{\pi(\bc)}{\pi'(\bc)}] = -kT \sum_i {c_i \ln
[\frac{\pi_i}{\pi'_i}]}.  \label{eq:Sippl1}
\end{equation}

Correspondingly, to calculate the effective potential energy
$\Delta H(\bc)$ of the system, one often assumes that $\Delta
H(\bc)$ can be decomposed into various basic energetic terms. For
a linear potential function,  $\Delta H(\bc)$ can be calculated
as:
\begin{equation}
\Delta H(\bc) = \sum_i \Delta H(c_i) = \sum_i c_i w_i.
\label{eq:Sippl2}
\end{equation}
If the distribution of each $c_i$ is assumed to be linearly
independent to the others in the native protein structures, we
have:
\begin{equation}
w_i  = -kT \ln [\frac{\pi_i}{\pi'_i}]. \label{Sippl}
\end{equation}
In another word, the probability of each structural feature in
native protein structures follows the Boltzmann distribution. This
is the {\it Boltzmann assumption\/} made in nearly all statistical
potential functions. Finkelstein (1995) summarized protein
structural features which are observed to correlate with the
Boltzmann distribution.  These include the distribution of
residues between the surface and interior of globules, the
occurrence of various $\phi,\psi,\chi$ angles, {\it cis} and {\it
trans} prolines, ion pairs, and empty cavities in protein
globules~\citep{whyBoltzmann}.

The probability $\pi_i$ can be estimated by counting frequency of
the $i$-th structural feature after combining all structures in
the database. Clearly, the probability $\pi_i$ is determined once
a database of crystal structures is given. The probability
$\pi'_i$ is calculated as the probability of the $i$-th structural
feature in the reference state. Therefore, the choice of the
reference state has large effects and is critical for developing
knowledge-based statistical potential function.

\subsection{Miyazawa-Jernigan contact potential}
Because of the importance of the Miyazawa-Jernigan model in
developing statistical knowledge-based potential and its wide use, we
discuss the Miyazawa-Jernigan contact potential in details.  This
also gives an exposure of different technical aspects of
developing statistical knowledge-based potential functions.

\vspace*{.15 in} \noindent {\bf Residue representation and contact
definition.} In the Miyazawa-Jernigan model, the $l$-th residue is
represented as single ball located at its side-chain center
$\bz_l$. If the $l$-th residue is a Gly residue, which lacks a
side chain, the positions of the C$^{\alpha}$ atom is taken as
$\bz_l$. A pair of residues $(l, m)$ are defined to be in contact
if the distance between their side-chain centers is less than a
threshold $\theta = 6.5$
 $\mathring{\mbox{A}}$. Neighboring residues $l$ and $m$ along amino
acid sequences ($|l-m|=1$) are excluded from statistical counting
because they are likely to be in spatial contact that does not
reflect the intrinsic preference for inter-residue interactions.
Thus, a contact between the $l$-th and $m$-th residues is defined
using $\Delta_{(l,m)}$:
$$
\Delta_{(l,m)} = \left\{ \begin{array}{ll}
1, & \mbox{if $|\bz_l-\bz_m| \le \theta$} \mbox{ and } |l-m| > 1; \\
0, & \mbox{otherwise}, \end{array} \right.
$$
where $| \bz_l - \bz_m|$ is the Euclidean distance between the
$l$-th and $m$-th residues.  Hence, the total number count of
$(i,j)$ contacts of residue type $i$ with residue type $j$ in
protein $p$ is:
\begin{equation}
n_{(i,j);\:p} = \sum_{\substack{l,m,\\ l < m}} \Delta_{(l,m)},
\quad \mbox{if ($\mathbb{I}(l), \mathbb{I}(m)) = (i,j)$ or
$(j,i)$}, \label{eq:1}
\end{equation}
where $\mathbb{I}(l)$ is the residue type of the $l$-th amino acid residue.
The total number count of $(i,j)$ contacts in all proteins are then:
\begin{equation}
n_{(i,j)} = \sum_p n_{(i,j);\:p}, \quad i, j = 1, 2, \cdots, 20.
\label{eq:2}
\end{equation}

\vspace*{.15 in} \noindent {\bf Coordination and solvent
assumption.} The number of different types of pairwise
residue-residue contacts $n_{(i,j)}$ can be counted directly from
the structure of proteins following Equation~\ref{eq:2}. We also
need to count the number of residue-solvent contacts. Since
solvent molecules are not consistently present in X-ray crystal
structures, and therefore cannot be counted exactly, Miyazawa and
Jernigan made an assumption based on the model of an effective
solvent molecule, which has the volume of the average volume of
the 20 types of residues. Physically, one effective solvent
molecule may represent several real water molecules or other
solvent molecules. The number of residue-solvent contacts
$n_{(i,0)}$ can be estimated as:
\begin{equation}
n_{(i,0)} = q_{i}n_{i} - (\sum^{20}_{\substack{j=1; \\ j \ne
i}}{n_{(i,j)}}+2n_{(i,i)}),
\label{eq:3}
\end{equation}
where the subscript $0$ represents the effective solvent molecule;
the other indice $i$ and $j$ represent the types of amino acids;
$n_{(i)}$ is the number of residue type $i$ in the set of
proteins; $q_i$ is the mean coordination number of buried residue
$i$, calculated as the number of contacts formed by a buried
residue of type $i$ averaged over a structure database. Here the
assumption is that residues make the same number of contacts on
average, with either effective solvent molecules (first term in
Equation~(\ref{eq:3}), or other residues (second term in
Equation~(\ref{eq:3})).

For convenience, we calculate the total numbers of residues
$n_{(r)}$, of residue-residue contacts $n_{(r,r)}$, of
residue-solvent contacts $n_{(r, 0)}$, and of pairwise contacts of
any type $n_{(\cdot,\cdot)}$ as follows:
\[
\begin{split}
n_{(r)} = \sum^{20}_{i = 1} n_{i}; \:\:\:\: n_{(i,r)} = n_{(r,i)}
= \sum^{20}_{j = 1} n_{(i,j)}; \:\:\: n_{(r,r)}
= \sum^{20}_{i = 1} n_{(i,r)}; \\
n_{(r,0)} = n_{(0,r)} = \sum^{20}_{i = 1}n_{(i,0)};\quad \quad
n_{(\cdot,\cdot)} = n_{(r,r)} + n_{(r,0)} + n_{(0,0)}.
\end{split}
\]

\begin{figure}[!t]
\centerline{\epsfig{figure=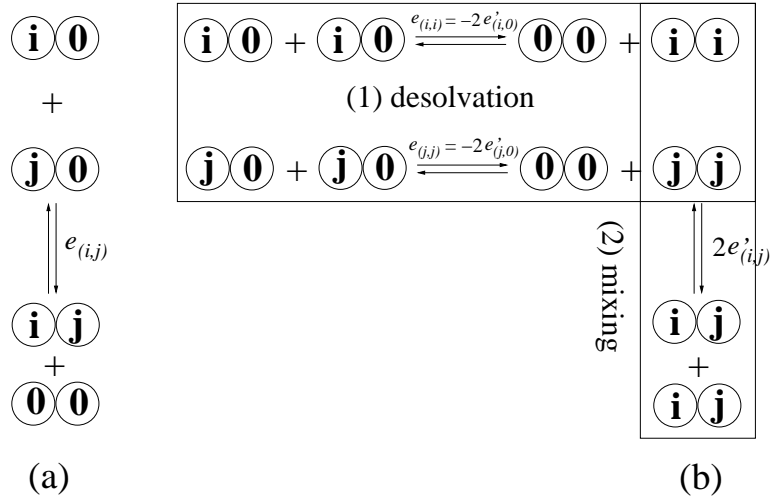,width=4in}}
\caption[test]{\small \sf The Miyazawa-Jernigan model of chemical
reaction. Amino acid residues first go through the desolvation
process, and then mix together to form pair contact interactions.
The associated free energies of desolvation $e_{(i,i)}$ and mixing
$e'_{(i,j)}$ can be obtained from the equilibrium constants of
these two processes. } \label{Fig:MJ}
\end{figure}

\vspace*{.15 in} \noindent {\bf Chemical reaction model.} Miyazawa
and Jernigan (1985) developed a physical model based on
hypothetical chemical reactions. In this model, residues of type
$i$ and $j$ in solution need to be desolvated before they can form
a contact.  The overall reaction is the formation of $(i,j)$
contacts, depicted in Figure~4.1a. The total free energy
change to form one pair of $(i,j)$ contact from fully solvated
residues of $i$ and $j$ is (Figure~4.1a):
\begin{equation}
e_{(i,j)} = (E_{(i,j)} + E_{(0,0)}) - (E_{(i,0)} + E_{(j,0)}), \label{eq:6a}\\
\end{equation}
where $E_{(i,j)}$ is the absolute contact energy between the
$i$-th and $j$-th types of residues, and $E_{(i,j)} =
E_{(j,\,i)}$; $E_{(i,0)}$ are the absolute contact energy between
the $i$-th residue and effective solvent, and $E_{(i,0)} =
E_{(0,i)}$; likewise for $E_{(j,\,0)}$; $E_{(0,0)}$ are the
absolute contact energies of solvent-solvent contacts $(0,0)$.

The overall reaction can be decomposed into two steps
(Figure~4.1b).  In the first step, residues of type $i$
and type $j$, initially fully solvated, are desolvated or
``demixed from solvent'' to form self-pairs $(i,i)$ and $(j,j)$.
The free energy changes $e_{(i,i)}$ and $e_{(j,j)}$ upon this
desolvation step can be easily seen from the desolvation process
(horizontal box) in Figure~4.1 as:
\begin{equation}
\setlength{\arraycolsep}{.04in}
\begin{array}{rclllll}
e_{(i,\,i)} &=& E_{(i,\,i)} &+& E_{(0,\,0)} &-& 2E_{(i,\,0)};\\
e_{(j,\,j)} &=& E_{(j,\,j)} &+& E_{(0,\,0)} &-& 2E_{(j,\,0)},
\label{eq:4}
\end{array}
\end{equation}
where $E_{(i,i)}, E_{(j,j)}$ are the absolute contact energies of
self pair $(i,i)$ and $(j,j)$, respectively.  In the second step,
the contacts in $(i,i)$ and $(j,j)$ pairs are broken and residues
of type $i$ and residues of type $j$ are mixed together to form
two $(i,j)$ pairs. The free energy change upon this mixing step
$2e'_{(i,j)}$ is (vertical box in Figure~4.1):
\begin{equation}
2e'_{(i,j)} = 2E_{(i,j)} - (E_{(i,i)} + E_{(j,j)}). \label{eq:5}
\end{equation}
Denote the free energy changes upon the mixing of residue of type
$i$ and solvent as $e'_{(i,0)}$, We have:
\begin{equation}
-2e'_{(i,0)} = e_{(i,i)} \quad \mbox{ and } -2e'_{(j,0)} = e_{(j,j)},
\end{equation}
which can be obtained from Equation~\ref{eq:4} and
Equation~\ref{eq:5} after substituting ``$j$'' with ``$0$''.
Following the reaction model of Figure~4.1b, the total
free energy change to form one pair of $(i,j)$ can be written as:
\begin{subequations}
\begin{eqnarray}
2e_{(i,j)}
&=& 2e'_{(i,j)} + e_{(i,i)}+e_{(j,j)} \label{eq:6b}\\
&=& 2e'_{(i,j)} - 2e'_{(i,0)} - 2e'_{(j,0)} \label{eq:6c}
\end{eqnarray}
\end{subequations}

\vspace*{.15 in} \noindent {\bf Contact energy model.} The total
energy of the system is due to the contacts between
residue-residue, residue-solvent, solvent-solvent:
\begin{equation}
\begin{split}
E_c &= \sum^{20}_{i = 0}\sum^{20}_{\substack {j = 0;\\j \ge i}}
{E_{(i,j)}n_{(i,j)}} \\
&= \sum^{20}_{i=1} \sum^{20}_{\substack{j=1;\\j \ge i}} E_{(i,j)}n_{(i,j)}
+ \sum^{20}_{i=1} E_{(i,0)}n_{(i,0)} + E_{(0,0)}n_{(0,0)}
\end{split}
\label{eq:7}
\end{equation}
Because the absolute contact energies $E_{(i,j)}$ is difficult to
measure and knowledge of this value is unnecessary for studying
the dependence of energy on protein conformation, we can simplify
Equation~\ref{eq:7} further. Our goal is to separate out terms
that do not depend on contact interactions and hence do not depend
on the conformation of the molecule. Equation~\ref{eq:7} can be
re-written as:
\begin{subequations}
\begin{eqnarray}
E_c &=& \sum^{20}_{i = 0}{(2E_{(i,0)} -
E_{(0,0)})q_in_{(i)}/2}. +\sum^{20}_{i = 1}\sum^{20}_{\substack{j = 1;\\
j \ge i}} {e_{(i,j)}n_{(i,j)}} \label{eq:8a}\\
&=& \sum^{20}_{i = 0}{E_{(i,i)}q_in_{(i)}/2} + \sum^{20}_{i =
0}\sum^{20}_{\substack{j = 0; \\j \ge i}} {e'_{(i,j)}n_{(i,j)}}
\label{eq:8b}
\end{eqnarray}
\end{subequations}
by using Equation~\ref{eq:3} and Equation~\ref{eq:6a}. Here only
the second terms in Equation~\ref{eq:8a} and \ref{eq:8b} are
dependent on protein conformations. Therefore, only either
$e_{(i,j)}$ or $e'_{(i,j)}$ needs to be estimated.  Since the
number of residue-residue contacts can be counted directly while
the number of residue-solvent contacts is more difficult to
obtain, Equation~\ref{eq:8a} is more convenient for calculating
the total contact energy of protein conformations.  Both
$e_{(i,j)}$ and $e'_{(i,j)}$ are termed as {\it effective contact
energies\/} and their values were reported in
\citep{Miyazawa&Jernigan96_JMB}.

\vspace*{.15 in} \noindent {\bf Estimating effective contact
energies: quasi-chemical approximation.} The effective contact
energies $e_{(i,j)}$ in Equation~\ref{eq:8a} can be estimated in
$kT$ unit by assuming that the solvent and solute molecules are in
quasi-chemical equilibrium for the reaction depicted in
Figure~4.1a:
\begin{equation}
e_{(i,j)} = -\ln{ \frac{[
m_{(i,j)}/m_{(\cdot,\cdot)}] [m_{(0,0)}/m_{(\cdot,\cdot)}]
}
{
[m_{(i,0)}/m_{(\cdot,\cdot)}] [m_{(j,0)}/m_{(\cdot,\cdot)}]
}
} 
= -\ln{ \frac{
m_{(i,j)} m_{(0,0)}
}
{
m_{(i,0)} m_{(j,0)}
}
} 
\label{eq:9}
\end{equation}
where $m_{(i,j)}, m_{(i,0)}$, and $m_{(0,0)}$ are the contact
numbers of pairs between residue type $i$ and $j$, residue type
$i$ and solvent, and solvent and solvent, respectively.
$m_{(\cdot,\cdot)}$ is the total number of contacts in the system
and is canceled out. Similarly, $e'_{(i,j)}$ and $e'_{(i,0)}$ can
be estimated from the model depicted in Figure~4.1b:
\begin{subequations}
\begin{eqnarray}
2e'_{(i,j)} &=& -\ln{\frac{[m_{(i,j)}]^2}{m_{(i,i)} m_{(j,j)}}};
\label{eq:10a}\\
2e'_{(i,0)} &=& -\ln{\frac{[m_{(i,0)}]^2}{m_{(i,i)} m_{(0,0)}}}
\label{eq:10b},
\end{eqnarray}
\label{eq:10}
\end{subequations}

Based on these models, two different techniques have been
developed to obtain effective contact energy parameters. Following
the hypothetical reaction in Figure~4.1(a), $e_{(i,j)}$
can be directly estimated from Equation~\ref{eq:9}, as was done by
Zhang and Kim~\citep{Zhang&Kim00_PNAS}. Alternatively, one can
follow the hypothetical two-step reaction in Figure~4.1b
and estimate each term in Equation~\ref{eq:6c} for $e_{(i,j)}$ by
using Equation~\ref{eq:10}. Because the second approach leads to
additional insight about the desolvation effects ($e'_{(i,0)}$)
and the mixing effects ($e'_{(i,j)}$) in contact interactions, we
follow this approach in subsequent discussions. The first approach
will become self-evident after our discussion.

\vspace*{.15 in} \noindent {\bf Models of  reference state.} In
reality, the true fraction $\frac{m_{(i,j)}} {m_{(\cdot, \cdot)}}$
of contacts of $(i,j)$ type among all pairwise contacts $(\cdot,
\cdot)$ is unknown. One can approximate this by calculating its
mean value from sampled structures in the database. We have:
\[
\frac{m_{(i,j)}} {m_{(\cdot, \cdot)}}
\approx \displaystyle\frac{\sum_p{n_{(i,j);p}}}{\sum_p{n_{(\cdot,
\cdot);p}}}; \quad \frac{m_{(i,0)}}{m_{(\cdot, \cdot)}} \approx
\displaystyle\frac{\sum_p{n_{(i,0);p}}}{\sum_p{n_{(\cdot,
\cdot);p}}}; \quad \frac{m_{(0,0)}}{m_{(\cdot, \cdot)}} \approx
\displaystyle\frac{\sum_p{n_{(0,0);p}}}{\sum_p{n_{(\cdot,
\cdot);p}}},
\]
where $i$ and $j \ne 0$. However, this yields a biased estimation
of $e'_{(i,j)}$ and $e_{(i,j)}$. When effective solvent molecules,
residues of $i$-th type and residues of $j$-th type are randomly
mixed, $e'_{(i,j)}$ will not equal to 0 as should be because of
differences in amino acid composition among proteins in the
database. Therefore, a reference state must be used to remove this
bias.

In the work of Miyazawa and Jernigan, the effective contact
energies for mixing two types of residues $e'_{(i,j)}$ and for
solvating a residue $e'_{(i,0)}$ are estimated based on two
different random mixture reference
states~\citep{Miyazawa&Jernigan85_M}. In both cases, the contacting
pairs in a structure are randomly permuted, but the global
conformation is retained. Hence, the total number of
residue-residue, residue-solvent, solvent-solvent contacts remain
unchanged.

The first random mixture reference state for desolvation contains
the same set of residues of the protein $p$ and a set of effective
solvent molecules. We denote the overall number of $(i,i), (i,0),
(0,0)$ contacts in this random mixture state after summing over
all proteins as $c'_{(i,i)}, c'_{(i,0)}$, and $c'_{(0,0)}$,
respectively. $c'_{(i,i)}$ can be computed as:
\begin{equation}
c'_{(i,i)} = \displaystyle\sum_p { [ \frac{q_i\:n_{i
;p}}{\displaystyle\sum_{k}{q_k\:n_{k;p}}} ] ^2 \cdot n_{(\cdot,
\cdot);p} },\label{eq:somelabel}
\end{equation}
where Miyazawa and Jernigan assumed that the average coordination
number of residue $i$ in all proteins is $q_i$. Therefore, a
residue of type $i$ makes $q_i n_{i;p}$ number of contacts in
protein $p$. Similarly, the number of $(i,0)$ contacts
$c'_{(i,0)}$ can be computed as:
\begin{equation}
c'_{(i,0)} = \sum_p\displaystyle{ [
\frac{q_i\:n_{i;p}}{\displaystyle\sum_{k}{q_k\:n_{k;p}}}} ] \;
n_{(\cdot,0);p}.
\end{equation}
From the horizontal box in Figure~4.1, the effective
contact energy $e'_{(i,0)}$ can now be computed as:
\begin{equation}
2e'_{(i,0)} = -\ln {\left[
\frac{n^2_{(i,0)}}{n_{(i,i)}n_{(0,0)}}\:/\:\frac{{c'\,^2_{(i,0)}}}{c'_{(i,i}c'_{(0,0)}}
\right]} \quad \quad \quad (i \ne 0). \label{eq:12}
\end{equation}

The second random mixture reference state for mixing contains the
exact same set of residues as the protein $p$, but have all
residues randomly mixed. We denote the number of $(i,j)$ contacts
in this random mixture as $c_{(i,j); p}$. The overall number of
$(i,j)$ contacts in the full protein set $c_{(i,j)}$ is the sum of
$c_{(i,j); p}$ over all proteins:
\begin{equation}
c_{(i,j)} = \displaystyle\sum_p { [\frac{n_{(i,
\cdot);p}}{n_{(\cdot, \cdot);p}}] [\frac{n_{(j,
\cdot);p}}{n_{(\cdot, \cdot);p}}] \cdot n_{(\cdot, \cdot);p} }.
\label{eq:13}
\end{equation}
From the vertical box in Figure~4.1, the effective
contact energy $e'_{(i,j)}$ can now be computed as:
\begin{equation}
2e'_{(i,j)} = -\ln {\left[
\frac{n^2_{(i,j)}}{n_{(i,i)}n_{(j,j)}}\:/\:\frac{c^2_{(i,j)}}{c_{(i,i)}c_{(j,j)}}
\right]}, \quad \quad \quad i \mbox{ or\,} j \ne 0. \label{eq:11}
\end{equation}
The compositional bias is removed by the denominator in
Equation~\ref{eq:11}, and $e'_{(i,j)}$ now equals to 0.

\begin{table}[!t]
\setlength{\tabcolsep}{.5pt}
\begin{tiny}
\begin{center}
\caption{Contact energies in $kT$ units; $e_{(i,j)}$ for upper
half and diagonal and $e'_{(i,j)}$ for lower half (from
\citep{Miyazawa&Jernigan96_JMB})} \label{Tab:MJ}
\begin{tabular}{l rrrrrrrrrrrrrrrrrrrr}
\hline \\&\scriptsize \texttt{Cys}&\scriptsize
\texttt{Met}&\scriptsize \texttt{Phe}&\scriptsize
\texttt{Ile}&\scriptsize \texttt{Leu}&\scriptsize
\texttt{Val}&\scriptsize \texttt{Trp}&\scriptsize
\texttt{Tyr}&\scriptsize \texttt{Ala}&\scriptsize \texttt{Gly}
&\scriptsize \texttt{Thr}&\scriptsize \texttt{Ser}&\scriptsize \texttt{Asn}&\scriptsize \texttt{Gln}&\scriptsize \texttt{Asp}&\scriptsize \texttt{Glu}&\scriptsize \texttt{His}&\scriptsize \texttt{Arg}&\scriptsize \texttt{Lys}&\scriptsize \texttt{Pro}\\
\hline \\
\scriptsize\texttt{Cys}&\underline{-5.44}&-4.99&-5.80&-5.50&-5.83&-4.96&-4.95&-4.16&-3.57&-3.16&-3.11&-2.86&-2.59&-2.85&-2.41&-2.27&-3.60&-2.57&-1.95&-3.07\\[1.5pt]
\scriptsize \texttt{Met}&0.46&\underline{-5.46}&-6.56&-6.02&-6.41&-5.32&-5.55&-4.91&-3.94&-3.39&-3.51&-3.03&-2.95&-3.30&-2.57&-2.89&-3.98&-3.12&-2.48&-3.45\\[1.5pt]
\scriptsize \texttt{Phe}&0.54&-0.20&\underline{-7.26}&-6.84&-7.28&-6.29&-6.16&-5.66&-4.81&-4.13&-4.28&-4.02&-3.75&-4.10&-3.48&-3.56&-4.77&-3.98&-3.36&-4.25\\[1.5pt]
\scriptsize \texttt{Ile}&0.49&-0.01&0.06&\underline{-6.54}&-7.04&-6.05&-5.78&-5.25&-4.58&-3.78&-4.03&-3.52&-3.24&-3.67&-3.17&-3.27&-4.14&-3.63&-3.01&-3.76\\[1.5pt]
\scriptsize \texttt{Leu}&0.57&0.01&0.03&-0.08&\underline{-7.37}&-6.48&-6.14&-5.67&-4.91&-4.16&-4.34&-3.92&-3.74&-4.04&-3.40&-3.59&-4.54&-4.03&-3.37&-4.20\\[1.5pt]
\scriptsize \texttt{Val}&0.52&0.18&0.10&-0.01&-0.04&\underline{-5.52}&-5.18&-4.62&-4.04&-3.38&-3.46&-3.05&-2.83&-3.07&-2.48&-2.67&-3.58&-3.07&-2.49&-3.32\\[1.5pt]
\scriptsize \texttt{Trp}&0.30&-0.29&0.00&0.02&0.08&0.11&\underline{-5.06}&-4.66&-3.82&-3.42&-3.22&-2.99&-3.07&-3.11&-2.84&-2.99&-3.98&-3.41&-2.69&-3.73\\[1.5pt]
\scriptsize \texttt{Tyr}&0.64&-0.10&0.05&0.11&0.10&0.23&-0.04&\underline{-4.17}&-3.36&-3.01&-3.01&-2.78&-2.76&-2.97&-2.76&-2.79&-3.52&-3.16&-2.60&-3.19\\[1.5pt]
\scriptsize \texttt{Ala}&0.51&0.15&0.17&0.05&0.13&0.08&0.07&0.09&\underline{-2.72}&-2.31&-2.32&-2.01&-1.84&-1.89&-1.70&-1.51&-2.41&-1.83&-1.31&-2.03\\[1.5pt]
\scriptsize \texttt{Gly}&0.68&0.46&0.62&0.62&0.65&0.51&0.24&0.20&0.18&\underline{-2.24}&-2.08&-1.82&-1.74&-1.66&-1.59&-1.22&-2.15&-1.72&-1.15&-1.87\\[1.5pt]
\scriptsize \texttt{Thr}&0.67&0.28&0.41&0.30&0.40&0.36&0.37&0.13&0.10&0.10&\underline{-2.12}&-1.96&-1.88&-1.90&-1.80&-1.74&-2.42&-1.90&-1.31&-1.90\\[1.5pt]
\scriptsize \texttt{Ser}&0.69&0.53&0.44&0.59&0.60&0.55&0.38&0.14&0.18&0.14&-0.06&\underline{-1.67}&-1.58&-1.49&-1.63&-1.48&-2.11&-1.62&-1.05&-1.57\\[1.5pt]
\scriptsize \texttt{Asn}&0.97&0.62&0.72&0.87&0.79&0.77&0.30&0.17&0.36&0.22&0.02&0.10&\underline{-1.68}&-1.71&-1.68&-1.51&-2.08&-1.64&-1.21&-1.53\\[1.5pt]
\scriptsize \texttt{Gln}&0.64&0.20&0.30&0.37&0.42&0.46&0.19&-0.12&0.24&0.24&-0.08&0.11&-0.10&\underline{-1.54}&-1.46&-1.42&-1.98&-1.80&-1.29&-1.73\\[1.5pt]
\scriptsize \texttt{Asp}&0.91&0.77&0.75&0.71&0.89&0.89&0.30&-0.07&0.26&0.13&-0.14&-0.19&-0.24&-0.09&\underline{-1.21}&-1.02&-2.32&-2.29&-1.68&-1.33\\[1.5pt]
\scriptsize \texttt{Glu}&0.91&0.30&0.52&0.46&0.55&0.55&0.00&-0.25&0.30&0.36&-0.22&-0.19&-0.21&-0.19&0.05&\underline{-0.91}&-2.15&-2.27&-1.80&-1.26\\[1.5pt]
\scriptsize \texttt{His}&0.65&0.28&0.39&0.66&0.67&0.70&0.08&0.09&0.47&0.50&0.16&0.26&0.29&0.31&-0.19&-0.16&\underline{-3.05}&-2.16&-1.35&-2.25\\[1.5pt]
\scriptsize \texttt{Arg}&0.93&0.38&0.42&0.41&0.43&0.47&-0.11&-0.30&0.30&0.18&-0.07&-0.01&-0.02&-0.26&-0.91&-1.04&0.14&\underline{-1.55}&-0.59&-1.70\\[1.5pt]
\scriptsize \texttt{Lys}&0.83&0.31&0.33&0.32&0.37&0.33&-0.10&-0.46&0.11&0.03&-0.19&-0.15&-0.30&-0.46&-1.01&-1.28&0.23&0.24&\underline{-0.12}&-0.97\\[1.5pt]
\scriptsize \texttt{Pro}&0.53&0.16&0.25&0.39&0.35&0.31&-0.33&-0.23&0.20&0.13&0.04&0.14&0.18&-0.08&0.14&0.07&0.15&-0.05&-0.04&\underline{-1.75}\\[1.5pt]
\hline
\end{tabular}
\end{center}
\end{tiny}
\end{table}

Although $c'_{(0,0)}$ can be estimated from Equation
(\ref{eq:10b}) by assuming that $e'_{(i,0)} = 0$ in a reference
state, Zhang and DeLisi (1997) simplified the Miyazawa-Jernigan
process by further assuming that the number of solvent-solvent
contacts in both reference states is the same as in the native
state~\citep{Zhang&Delisi97_JMB}:
\begin{equation}
c'_{(0,0)}= n_{(0,0)}. \label{eq:16}
\end{equation}
Therefore, $c'_{(0,0)}$ and $n_{(0,0)}$ are canceled out in
Equation~\ref{eq:12} and not needed for calculating $e'_{(i,0)}$.
This treatment systematically subtracts a constant scaling energy
from all effective energies $e_{(i,j)}$, and should produce
exactly the same relative energy values for protein conformations
as Miyazawa-Jernigan's original work, with the difference of a
constant offset value. In fact, Miyazawa and Jernigan (1996)
showed that this constant scaling energy is the effective contact
energy $e_{\hat{r}\hat{r}}$ between the average residue $\hat{r}$
of the 20 residue types, and suggested that
$e_{(i,j)}-e_{\hat{r}\hat{r}}$ being used to measure the stability
of a protein structure~\citep{Miyazawa&Jernigan96_JMB}.

\vspace*{.15 in} \noindent {\bf Hydrophobic nature of
Miyazawa-Jernigan contact potential.} In the relation of
Equation~\ref{eq:6c}, $e_{(i,j)}= e'_{(i,j)} - (e'_{(i,0)} +
e'_{(j,0)})$, the Miyazawa-Jernigan effective contact energy
$e_{(i,j)}$ is composed of two types of terms: the desolvation
terms $e'_{(i,0)}$ and $e'_{(j,0)}$ and the mixing term
$e'_{(i,j)}$. The desolvation term of residue type $i$, that is,
$-e'_{(i,0)}$ or $e_{(i,i)}/2$ (Figure~4.1), is the
energy change due to the desolvation of residue $i$, the formation
of the $i$-$i$ self-pair, and the solvent-solvent pair. The value
of this term $e_{(i,i)}/2$ should correlate well with the
hydrophobicity of residue type $i$
\citep{Miyazawa&Jernigan85_M,Li97_PRL}, although for charged amino
acids this term also incorporates unfavorable electrostatic
potentials of self-pairing. The mixing term $e'_{(i,j)}$ is the
energy change accompanying the mixing of two different types of
amino acids of $i$ and $j$ to form a contact pair $i$-$j$ after
breaking self-pairs $i$-$i$ and $j$-$j$. Its value measures the
tendency of different residues to mix together. For example, the
mixing between two residues with opposite charges are more
favorable than mixing between other types of residues, because of
the favorable electrostatic interactions.

Important insights into the nature of residue-residue contact
interactions can also be obtained by a quantitative analysis of
the desolvation terms and the mixing terms. Among different types
of contacts, the average difference of the desolvation terms is 9
times larger than that of the mixing terms (see Table~\ref{Tab:MJ}
taken from \citep{Miyazawa&Jernigan96_JMB}). Thus, a comparison of
the values of $(e_{(i,i)}+e_{jj})/2$ and $e'_{(i,j)}$ clearly
shows that the desolvation term plays the dominant role in
determining the energy difference among different conformations.

Similar conclusion can be drawn by an eigenvalue decomposition
analysis of the Miyazawa-Jernigan matrix $\bM$, which is made up
of $e_{(i,j)}$ values alone, without the knowledge of the mixing
terms $e'_{(i,j)}$~\citep{Li97_PRL}. The $\bM$ matrix is a $20
\times 20$ real symmetric matrix, and thus can be reconstructed
based on the following spectral decomposition:
\begin{equation}
e_{(i,j)} = [\sum^N_{k = 1} \lambda_k \bv_k \bv_k]_{ij}=\sum^N_{k
= 1} \lambda_k \bv_k(i) \bv_k(j), \label{eq:17}
\end{equation}
where $\lambda_k$ and $\bv_k$ is the $k$-th largest eigenvalue and
the corresponding eigenvector, respectively; $\bv_k(i)$ is the
$i$-th component of the $k$-th eigenvector. Li {\it et al.\/}
(1997) found that there are two dominant eigenvalues $\lambda_1$
and $\lambda_2$, and the corresponding two eigenvectors are
strongly correlated after a shift and a rescaling operation, {\it
i.e.}, $\bv_2 = \alpha \bu + \beta \bv_1$. Here, $\bu$ is the
${\bf 1}$ vector with each component equals to 1 and $\alpha$ and
$\beta$ are scalars. Therefore, $\bM$ can be well-approximated
with only one eigenvector $\bv_1$ corresponding to the largest
eigenvalue $\lambda_1$. For each entry $e_{(i,j)}$ of the matrix
$\bM$, we have the following approximation:
\begin{equation}
e_{(i,j)}  \approx \lambda_1 \bv_1(i) \bv_1(j) + \lambda_2
\bv_2(i)
\bv_2(j)
\approx c_0 + c_1(q_i + q_j) + c_2q_iq_j, \label{eq:18}
\end{equation}
where $q_i \equiv v_1(i)$, and $c_0, c_1$ and $c_2$ are constants.
To better understand the underlying physical implications,
Equation~\ref{eq:18} can be rewritten into the following form:
\begin{equation}
e_{(i,j)} \approx h_i + h_j - c_2(q_i - q_j)^2/2, \label{eq:19}
\end{equation}
where $$h_i = c_0/2 + c_1 q_i + (c_2/2)q^2_i.$$ Here $h_i + h_j$
is a single-body term and are interpreted as the desolvation term
in \citep{Li97_PRL}; $-c_2(q_i - q_j)^2/2$ is a two-body term and
are interpreted as the mixing term and the magnitude of the mixing
term is significantly smaller than that of $h_i + h_j$. This
result is not surprising and is consistent with the original model
of Miyazawa-Jernigan contact matrix $\bM$, where $e_{(i,j)} \equiv
e'_{(i,j)} - (e'_{(i,0)} +e'_{(j,0)}) $.

To summarize, the quantitative analysis of Miyazawa-Jernigan
contact energies reveals that hydrophobic effect is the dominant
driving force for protein folding.
To a large extent, this conclusion justifies the HP model proposed by
Chan and Dill (1990) where only hydrophobic interactions are included
in studies of simple models of protein folding~\citep{HP}.

\subsection{Distance dependent potential function}
In the Miyazawa-Jernigan potential function, interactions between
amino acids are assumed to be short-ranged and a distance cutoff
is used to define the occurrence of a contact. This type of
statistical potential is referred to as the ``contact potential''.
Another type of statistical potential allows modeling of residue
interactions that are distance-dependent. The distance of
interactions are usually divided into a number of small intervals
or bins, and the potential functions are derived by applying
Equation~\ref{Sippl} for individual distance intervals.

\vspace*{.25in} \noindent{\bf Formulation of distance-dependent
potential functions.} In distance-dependent statistical potential
functions, Equation~\ref{Sippl} can be written in several forms.
To follow the conventional notations, we use $(i,j)$ to represent
the $k$-th protein descriptor $c_k$ for pairwise interactions
between residue type $i$ and residue type $j$. From
Equation~\ref{Sippl}, we have:
\begin{subequations}
\begin{equation}
\begin{split}
\Delta H(i,j;\,d) &= - \ln \frac{\pi(i,j;\,d)}{\pi'(i,j;\,d)} = -
\ln \frac{n_{(i,j;\,d)}/n}{\pi'(i,j;\,d)} \\
&= - \ln \frac{n_{(i,j;\,d)}}{n'_{(i, j;\, d)}},
\label{distance_a}
\end{split}
\end{equation}
where $(i,j;\,d)$ represents an interaction between a specific
residue pair $(i,j)$ at distance $d$, $\Delta H(i,j;\,d)$ is the
the contribution from the $(i,j)$ type of residue pairs at
distance $d$, $\pi(\i,j;\,d)$ and $\pi'(i,j;\,d)$ are the observed
and expected probabilities of this distance-dependent interaction,
respectively, $n_{(i,j;\,d)}$ the observed number of $(i,j;\,d)$
interactions, $n$ the observed total number of all pairwise
interactions in a database, $n'_{(i,j;\,d)}$ the expected number
of $(\i, j;\,d)$ interactions when the total number of all
pairwise interactions in reference state is set to be $n$.

Since the expected joint probability $\pi'(i,j;\,d)$ for the
reference is not easy to estimate, Sippl (1990) replaces
Equation~\ref{Sippl} with:
\begin{equation}
\begin{split}
\Delta H(i,j;\,d) &= - \ln \frac{\pi(i,j\;|\;d)}{\pi'(i,j\;|\;d)}
= -
\ln \frac{n_{(i,j;\,d)}/n_{(d)}}{\pi'(i,j\;|\;d)} \\
&= - \ln \frac{n_{(i,j;\,d)}}{n'_{(i,j;\,d)}}, \label{distance_b}
\end{split}
\end{equation}
where $\pi(i,j\;|\;d)$ and $\pi'(i,j\;|\;d)$ are the observed and
expected probability of interaction of residue pairs $(i,j)$ given
the distance interval $d$, respectively; $n_{(d)}$ is the observed
total number of all pairwise interactions at the distance $d$;
$n'_{(i,j;\,d)} = \pi'(i,j\;|\;d) \cdot n(d)$ is the expected
number of $(i,j)$ interactions at $d$ when the total number of all
pairwise interactions at this distance $d$ in the reference state
is set to $n_{(d)}$. There are several variations of potential
function of this form, including the {\it ``Knowledge-Based
Potential function''} (KBP) by Lu and Skolnick
(2001)~\citep{Lu&Skolnick01_Proteins}.

In the work of developing the {\it ``Residue-specific All-atom
Probability Discriminatory Function''}
(RAPDF)~\citep{Samudrala&Moult98_JMB}, Samudrala and Moult (1998)
alternatively replaced Equation~\ref{Sippl} with:
\begin{equation}
\begin{split}
\Delta H(i,j;\,d) &= - \ln \frac{\pi(d\;|\;i,j)}{\pi'(d\;|\;i,j)}
= -
\ln \frac{n_{(i,j;\,d)}/n_{(i,j)}}{\pi'(d\;|\;i,j)}\\
&= - \ln \frac{n_{(i,j;\,d)}}{n'_{(i,j;\,d)}},\label{distance_c}
\end{split}
\end{equation}
\end{subequations}
where $\pi(d\;|\;i,j)$ and $\pi'(d\;|\;i,j)$ are the observed and
expected probability of interaction at the distance $d$ for a
given pair of residues $(i,j)$, respectively; $n_{(i,j)}$ is the
observed total number of interactions for $(i,j)$ pairs regardless
of the distance. $n'_{(i,j;\,d)} = \pi'(d\;|\;i,j) \cdot
n_{(i,j)}$ is the expected number of $(i,j)$ interactions at
distance $d$ when the total number of $(i,j)$ interactions in the
reference state is set to $n_{(d)}$.

The knowledge-based potential functions of Equation~\ref{distance_a},
~\ref{distance_b}, and ~\ref{distance_c} can all be written using
the unifying formula based on the number counts of interactions:
\begin{equation}
\Delta H(i,j;\,d) = - \ln [\frac{n_{(i,j;\,d)}}{n'_{(i,j;\,d)}}].
\label{distance_final}
\end{equation}
Clearly, the different ways of assigning $n'_{(i,j;\,d)}$ make the
potential functions differ from each other substantially, since
the method to calculate $n_{(i,j;\,d)}$ is essentially the same
for many potential functions. In other words, the model of
reference state used to compute $n'_{(i,j;\,d)}$ is critical for
distance-dependent energy functions.

\vspace*{.25in} \noindent{\bf Different models of reference
states.} Sippl (1990) first proposed the ``uniform density'' model
of reference state, where the probability density function for a
pair of contacting residues $(i,j)$ is uniformly distributed along
the distance vector connecting them: $\pi'(i,j\;|\;d) = \pi'(i,j)
$~\citep{Sippl90_JMB}. Lu and Skolnick made use of this type of
reference state to calculate the expected number of $(i,j)$
interactions at distance $d$ as~\citep{Lu&Skolnick01_Proteins}:
\[
n'_{(i,j;\,d)} = \pi'(i,j\;|\;d) \cdot n_{(d)} = \pi'(i,j) \cdot
n_{(d)}.
\]
The expected probability $\pi'(i,j)$ is estimated using the random
mixture approximation as:
\[
\pi'(i,j) = \chi_i \chi_j,
\]
where $\chi_i$ and $\chi_j$ are the mole fractions of residue type
$i$ and $j$, respectively.

Samudrala and Moult (1998) made use of another type of reference
state, where the probability of the distance between a pair of
residues $(i,j)$ being $d$ is independent of the contact types
$(i,j)$~\citep{Samudrala&Moult98_JMB}:
$$\pi'(d\;|\;i,j) = \pi'(d).
$$
The expected number of $(i,j)$ interactions at distance $d$ in
Equation~\ref{distance_c} becomes:
\[
n'_{(i,j;\,d)} = \pi'(d\;|\;i,j) \cdot n_{(i,j)} = \pi'(d) \cdot
n_{(i,j)},
\]
where $\pi'(r)$ is estimated from $\pi(r)$:
\[
\pi'(d) = \pi(d) = n_{(d)}/n.
\]

\vspace*{.15 in} \noindent {\bf Ideal gas reference state.}  In
the uniform density model of Sippl, the same density of a
particular residue pair $(i,j)$ along a line could result from
very different volume distribution of $(i,j)$ pairs in specific
regions of the protein.  For example, one spherical shell proximal
to the molecular center could be sparsely populated with residues,
and another distant shell could be densely populated, but all may
have the same density of $(i,j)$ pairs along the same radial
vector.  Zhou and Zhou (2002) developed a new reference state
(called {\sc Dfire} for ``Distance-scaled, Finite Ideal-gas
REference state'') where residues follow uniform distribution
everywhere in the protein~\citep{Zhou02_ProSci}. Assuming that
residues can be modeled as noninteracting points ({\it i.e.}, as
ideal gas molecules), the distribution of interacting pairs should
follow the uniform distribution not only along any vector lines,
but also in the whole volume of the protein.

When the distance between a pair of residues $(i,j)$ is at a
threshold distance $d_{\theta} = 14.5 $ $\mathring{\mbox{A}}$, the
interaction energy between them can be considered to be 0.
Therefore, residue type $i$ and type $j$ form pairs at the
distance $d_\theta$ purely by random, and the observed number of
$(i,j)$ pairs at the distance $d_\theta$ can be considered the
same as the expected number of $(i,j)$ pairs at the distance
$d_\theta$ in the reference state. Denote $v_d$ as the volume of a
spherical shell of width $\Delta d$ at a distance $d$ from the
center. The expected number of interactions $(i,j)$ at the
distance $d$ after volume correction is:
\[
n'_{(i,j;\,d)} = n_{(i,j;\,d_{\theta})} \cdot \frac{v_d }
{d_{\theta}} = n_{(i,j,d_{\theta})} \cdot
(\frac{d}{d_{\theta}})^{\alpha}\frac{\Delta d}{\Delta d_{\theta}}.
\]

For a protein molecule, $n'_{(i,j;\,d)}$ will not increase as
$r^2$ because of its finite size.  In addition, it is well-known
that the volume of protein molecule cannot be treated as a solid
body, as there are numerous voids and pockets in the interior.
This implies that the number density for a very large molecule
will also not scale as $d^2$ \citep{LiangDill01_BJ}.  Zhou and Zhou
(2002) assumed that $n'_{(i,j;\,d)}$ increase in $d^{\alpha}$
rather than $d^2$, where the exponent $\alpha$ needs to be
determined.  To estimate the $\alpha$ value, each protein $p$ in
the database is reshaped into a ball of radius $c_pR_{g;\,p}$,
where $R_{g;\,p}$ is the radius of gyration of the protein $p$,
and residues are distributed uniformly in this reshaped ball. Here
$c_p$ takes the value so that in the reshaped molecule, the number
of total interacting pairs at $d_\theta$ distance is about the
same as that observed in the native protein $p$, namely:
$$
\sum_{(i,j)} n'_{(i,j;\,d_\theta)} = \sum_{(i,j)}
n_{(i,j;\,d_\theta)}
$$
for protein $p$.
Once the value of $c_p$ is determined and hence the effective
radius $c_pR_{g;\,p}$ for each native protein is known, the number
of interacting pairs $n_{(d)}$ at distance $d$ can be counted
directly from the reshaped ball.  Zhou and Zhou further defined a
reduced distance-dependent function $f(d) = n_{(d)}/d^{\alpha}$
and the relative fluctuation $\delta$ of $f(d)$:
\[
\delta = [\frac{1}{n_b}\sum_d{(f(d) - \bar f)^2 / (\bar
f)}]^{1/2},
\]
where $\bar f = \sum_d{f(d)/n_b}$, and $n_b$ is the total number
of distance shells, all of which has the same thickness.  $\alpha$
is then estimated by minimizing the relative fluctuation $\delta$.
The rationale is that since idealized residues are points and are
uniformly distributed in the reshaped ball, $\delta$ should be
$0$. In their study, $\alpha$ was found to be $1.61$
~\citep{Zhou02_ProSci}.

\subsection{Geometric potential functions.} \label{subsec:geom}
The effectiveness of potential function also depends on the
representation of protein structures. Another class of knowledge-based
statistical potentials is based on the computation of various
geometric constructs that reflect the shape of the protein
molecules more accurately.  These geometric constructs include the
Voronoi diagram \citep{McConkey03_PNAS}, the Delaunay triangulation
\citep{Singh&Tropsha96_JCB,Zheng&Tropsha97_PSB,Tropsha01_JMB,Tropsha03_Bioinformatics},
and the alpha shape
\citep{Li&Liang03_Proteins,Li&Liang05_Proteins,Li&Liang05_PSB} of
the protein molecules.  Geometric potential functions has achieved
significant successes in many fields.  For example, the potential
function developed by McConkey {\it et al.\/} is based on the
Voronoi diagram of the atomic structures of proteins, and is among
one of the best performing atom-level potential functions in decoy
discrimination \citep{McConkey03_PNAS}.  Because the alpha shape of
the molecule contains rich topological, combinatorial, and metric
information, and has a strong theoretical foundation, we discuss
the alpha potential functions in more detail below as an example
of this class of potential function.

\vspace*{.15 in} \noindent {\bf Geometric model. }  In
Miyazawa-Jernigan and other contact potential functions, pairwise
contact interactions are declared if two residues or atoms are
within a specific cut-off distance.  Contacts by distance cut-off
can potentially include many implausible non-contacting neighbors,
which have no significant physical interaction
\citep{Smith99_Proteins}.  Whether or not a pair of residues can
make physical contact depends not only on the distance between
their center positions (such as C$_\alpha$ or C$_\beta$, or
geometric centers of side chain), but also on the size and the
orientations of side-chains \citep{Smith99_Proteins}. Furthermore,
two atoms close to each other may in fact be shielded from contact
by other atoms.  By occupying the intervening space, other
residues can block a pair of residues from direct interacting with
each other. Inclusion of these fictitious contact interactions
would be undesirable.

The alpha potential solves this problem by identifying interacting
residue pairs following the edges computed in the alpha shape.
Details of alpha shape can be found in the Chapter ``{\it Protein
structure geometry}''. When the parameter $\alpha$ is set to be 0,
residue contact occurs if residues or atoms from non-bonded
residues share a Voronoi edge, and this edge is at least partially
contained in the body of the molecule.  Figure~4.2
illustrates the basic ideas.

\begin{figure}[!t]
\centerline{\epsfig{figure=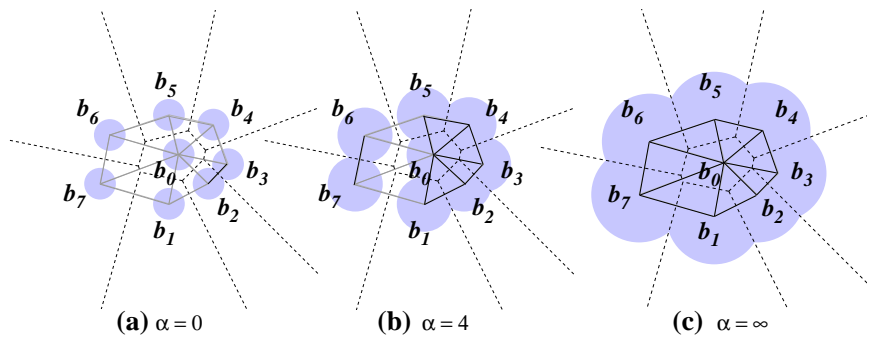,width=4.5in}} \caption
[test2]{\small \sf Schematic drawing of the Delaunay complex and
the alpha shape of a two-dimensional molecule. The Voronoi region
of a ball is the set of points closest to it when measured in
power distance. If two Voronoi regions share a boundary, {\it
i.e.}\ if there is a Voronoi edge (dashed line), we draw a
Delaunay edge (solid line in grey or black) between these two
Voronoi vertices. A Delaunay edge is therefore the {\it dual\/} of
a Voronoi edge. All Delaunay edges incident to ball residue $b_i$
form the {\it 1-star\/} for $b_i$, denoted as $St_1(b_i)$. When
the balls are inflated by increasing the $\alpha$ value, more
balls overlap, and more Voronoi edges intersect with the balls.
Therefore, more {\it dual\/} Delaunay edges are included in the
alpha shape (shown as black solid line segments). (a) When $\alpha
= 0.0$, the balls are not inflated and there is only one alpha
edge $\sigma_{_{2,3}}$ between ball $b_2$ and ball $b_3$; (b) When
$\alpha = 4.0$, the balls are inflated and their radii are
$\sqrt{r^2+4.0}$. There are six alpha edges: $\sigma_{_{0,1}},
\sigma_{_{0,2}}, \sigma_{_{0,3}}, \sigma_{_{0,4}},
\sigma_{_{0,4}}, \sigma_{_{0,5}},$ and $\sigma_{_{6,7}}$.  For a
ball $b_i$, the set of residue balls connected to it by alpha
edges are called the near neighbors of the ball. The number of
this set of residue balls is defined as the {\it degree of near
neighbors\/} of the residue ball $b_i$, denoted as $\rho_i$. For
example, $\rho_{_0} = 5$, and $\rho_{_7} =1$; (c) When $\alpha =
\infty$, all the Delaunay edges become alpha edges ($\alpha =
16.0$ is used for drawing). Hence, all long-range interactions not
intervened by a third residue are included.} \label{Fig:alpha}
\vspace*{2mm}
\end{figure}

\vspace*{.15 in} \noindent {\bf Distance and packing dependent
alpha potential.} For two non-bonded residue balls $b_i$ of radius
$r_i$ with its center located at $\bz_i$ and $b_j$ of radius $r_j$
at $\bz_j$, they form an alpha contact $(i,j\;|\;\alpha)$ if their
Voronoi regions intersect and these residue balls also intersect
after their radii are inflated to $r_i(\alpha) =
(r_i^2+\alpha)^{1/2}$ and $r_j(\alpha) = (r_j^2+\alpha)^{1/2}$,
respectively. That is, the alpha contact  $(i,j\;|\;\alpha)$
exists when:
\[
|\bz_i-\bz_j| < ({r_i}^2+\alpha)^{1/2}+({r_j}^2 + \alpha)^{1/2},
\quad \sigma_{i,j} \in {\cal K}_\alpha  \mbox{ and } |i-j| > 1.
\]

We further define the {\it 1-star\/} for each residue ball $b_i$ as:
$St_1 (b_i) = \{(b_i, b_j) \in \mathcal{K_\alpha}$, namely, the set of
1-simplices with $b_i$ as a vertex. The {\it near neighbors\/} of
$b_i$ are derived from $St_1(b_i)$ and are defined as:
\[
{\cal N}_\alpha(b_i) \equiv \{ b_j| \sigma_{i,j} \in
\mathcal{K}_\alpha \},  \quad \alpha = 4.0.
\]
and the {\it degree of near neighbors\/} $\rho_i$ of residue $b_i$ is
defined as the size of this set of residues:
\[
\rho_i \equiv |{\cal N}_\alpha(b_i)|,  \quad \alpha = 4.0.
\]
The degree of near neighbors $\rho_i$ is a parameter related to the
local packing density and hence indirectly the solvent accessibility
around the residue ball $b_i$ (Figure~4.2b). A large
$\rho_i$ value indicates high local packing density and less solvent
accessibility, and a small $\rho_i$ value indicates low local packing
density and high solvent accessibility.  Similarly, the {\it degree of
near neighbors\/} for a pair of residues is defined as:
\[
\rho_{(i,j)} \equiv |{\cal N}_\alpha(b_i,b_j)| = |{\cal
N}_\alpha(b_i)| + |{\cal N}_\alpha(b_j)|,  \quad \alpha = 4.0.
\]

\begin{figure}[!t]
\centerline{\epsfig{figure=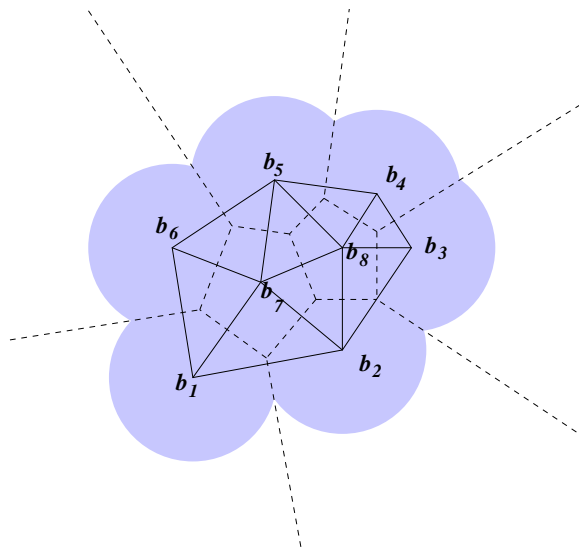,width=3in}}
\caption{\small \sf Non-interacting pairs. $(b_1,b_4)$ is
considered as a non-interacting pair because the shortest length
$L_{(1,4)}$ is equal to three, {\it i.e.}, the interaction between
$b_1$ and $b_4$ is blocked by two residues $b_7$ and $b_8$.
Likewise, $(b_3,b_6)$ is considered as a non-interacting pair as
well.} \label{Fig:reference}
\end{figure}

\noindent {\bf Reference state and collection of non-interacting
pairs.} We denote the shortest path length between residue $b_i$ and
residue $b_j$ as $L_{(i,j)}$, which is the fewest number of alpha
edges ($\alpha = 4)$ that connects $b_i$ and $b_j$.  The reference
state of the alpha potential is based on the collection of all
non-interacting residue pairs $(i,j)$:
         $$\{(i,j)|L_{(i,j)} = 3\};$$ Any $(i,j)$ pair in this
reference state is intercepted by two residues (Figure~4.3). We assume
that there is no attractive or repulsive interactions between them,
because of the shielding effect by the two intervening residues.
Namely, residue $i$ and residue $j$ form a pair only by random chance,
and any properties associated with $b_i$, such as packing density,
side-chain orientation, are independent of the same properties
associated with $b_j$.

\vspace*{.15 in} \noindent {\bf Statistical model: pairwise
potential and desolvation potential.} According to
Equation~(\ref{Sippl}, the packing and distance-dependent
statistical potential of residue pair $(k,l)$ at the packing
environment $\rho_{{(k,l)}}$ and the distance specified by
$\alpha$ is:

\begin{equation} H(k,l,\rho_{{(k,l)}}\;|\;\alpha) =
-KT \ln
(\frac{\pi_{(k,l,\,\rho_{{(k,l)}}\;|\;\alpha)}}{\pi'_{(k,l,\,\rho_{{(k,l)}})}}).
\end{equation}
Here, $\pi_{(k,l,\,\rho_{{(k,l)}}\;|\;\alpha)}$ is the observed
probability:
\begin{equation}
\pi_{(k,l,\,\rho_{{(k,l)}}\;|\;\alpha)} =
\frac{n_{(k,l,\,\rho_{{(k,l)}},\alpha)}} {n_{(\alpha)}},
\end{equation}
where $n_{(k,l,\,\rho_{{(k,l)}},\alpha)}$ is the number of residue
pair $(k,l)$ at the packing environment $\rho_{{(k,l)}}$ and the
distance specified by $\alpha$, and $n_{(\alpha)}$ is the total
number of residue pairs at the distance specified by $\alpha$.
$\pi'_{(k,l,\,\rho_{{(k,l)}})}$ is the expected probability:
\begin{equation}
\pi'_{(k,l,\,\rho_{{(k,l)}})} = \frac{n'_{(k,l,\,\rho_{(k,l)})}}
{n'},
\end{equation}
where $n'_{(k,l,z_{(k,l)})}$ is the number of residue pair $(k,l)$
at the packing environment $z_{(k,l)}$ in reference state, and
$n'$ is the total number of non-interacting residue pairs at the
reference state.

The desolvation potential of residue type $k$ to have $\rho$ near
neighbors $H(z\;|\;k)$ is estimated simply by following
Equation~(\ref{Sippl},:
\begin{equation}
H(\rho\;|\;k) = \frac{\pi_{(\rho\;|\;k)}}{\pi'_{(\rho\;|\;k)}} =
\frac{[n_{(k,\rho)}/n_{(k)}]}{[n_{(r,\rho)}/n_{(r)}]},
\end{equation}
where $r$ represent all 20 residue types.

For a protein structure, the total internal energy is estimated by
the summation of the desolvation energy and pairwise interaction
energy in the particular desolvated environment:
\begin{equation}
\begin{split}
H(s,a) & = \sum_{k,\rho} {H(\rho\;|\;k) \cdot n_{(k,\rho)}}\\
       & + \frac{1}{2}\sum_{k,l,\rho_{k,l},\alpha}{H(k,l,\rho_{(k,l)}\;|\;\alpha)
       \cdot n_{(k,l,\rho_{(k,l)},\alpha)}}
\end{split}
\end{equation}

\subsection{Sampling weight of proteins in database}
When developing statistical energy functions using a database
consisting of many homologous sequences, undesirable sampling
biases will be introduced. An easy way to avoid such sampling bias
is to construct a database of structures in which no pair of
proteins can have more than 25\% sequence identity. By this
criterion, a structure database may exclude a significant number
of informative structures, which may be valuable for studying a
specific type of proteins with very few known structures. An
alternative method to avoid such sampling bias without neglecting
these structures is to introduce weights that are properly
adjusted for each structure, which may or may not be homologous to
other structures in the database.

A similarity matrix $\bS$ of all proteins in the database can be
used to decide the weight for each protein structure
\citep{Miyazawa&Jernigan96_JMB}. The similarity between the $k$-th
and $l$-th proteins is defined by Miyazawa and Jernigan based on
the result of sequence alignment:
\begin{eqnarray*}
&\displaystyle{s_{kl} \equiv \frac{2\theta_{kl}}{L_k + L_l}}& \\
&\displaystyle{0 \le s_{kl} = s_{lk} \le 1}& \\
&\displaystyle{s_{kk} = 1}&
\end{eqnarray*}
where $\theta_{kl}$ is the number of identical residues in the
alignment, $L_k$ and $L_l$ are the lengths of sequences $k$ and
$l$, respectively. This similarity matrix {$\bS$} is symmetric and
composed of real values. It has the spectral decomposition:
\begin{equation}
\bS
= \sum_i \lambda_i {\bv_i} {\bv^T_i},
\end{equation}
where $\lambda_i$ and $\bv_i$ are the $i$-th eigenvalue and
eigenvector of $\bS$, respectively. For symmetric matrix, these
eigenvectors form an orthonormal base. Because for the symmetric
matrix $\bS$, $\sum_i\lambda_i =$ Trace($\bS$) =~
$n_{\mbox{prot}}$ and $\bS$ is positive semi-definite, we have:
\begin{equation}
0 \le \lambda_i \le n_{\rm prot},
\end{equation}
where $n_{\rm prot}$ is the number of proteins included in the
database. The value of $\lambda_i$ reflects the weight of the
corresponding orthogonal eigenvector $\bv_i$ to the matrix $\bS$.
For the special case where there one distinct sequence, which is
completely dissimilar to any other $n_{\rm prot} -1$ sequences in
the database, at least one eigenvalues will be exactly equal to
$1$ and the corresponding eigenvector represents this distinct
sequence but contains no information about other sequences due to
the orthogonality of the eigenvectors of matrix $\bS$. In another
case when there is one set of $m$ sequences which are exactly the
same within the group but are completely dissimilar to any other
$n_{\rm prot} -m$ sequences outside this set, at least one
eigenvalue will be exactly equal to $m$ and $m-1$ eigenvalues will
be equal to zero. The eigenvector corresponding to the non-zero
eigenvalue represents the whole group of those $m$ sequences but
contains no information about other sequences.

On the basis of these characteristics, Miyazawa and Jernigan
(1996) decreased all eigenvalues $>$ 1 to 1 to reconstruct a new
weight matrix $\bS'$, so that redundant information from similar
sequences are removed and the weight $w_k$ for the $k$th protein
in the database is determined. In another word, we have before
weighting:
\begin{equation}
w_k \equiv s_{kk} = \left[\sum_i \lambda_i {\bv_i}
{\bv^T_i}\right]_{kk} = 1\;
\end{equation}
after weighting,
\begin{equation}
w_k \equiv s'_{kk} = \left[\sum_i \lambda'_i {\bv_i}
{\bv^T_i}\right]_{kk}
\end{equation}
where
\[
\lambda'_i = \left\{\begin{array}{ll} \lambda_i, &\mbox{ if }
\lambda_i \le 1;\\
1, & \mbox{ if } \lambda_i > 1.
\end{array}\right.
\]
Therefore, if and only if a sequence is completely dissimilar to
any other sequences ($ \lambda'_i = \lambda_i = 1$), the sampling
weight for that sequence will be $1$. If all $n_{prot}$ sequences
in the database are identical, the sampling weights for these
sequences will be  $1/n_{\rm prot}$. Generally, sampling weights
take a value between one and $1/n_{\mbox{prot}}$, and are about
negatively proportional to the number of similar sequences.

\section{Optimization method}
There are several drawbacks of knowledge-based potential function
derived from statistical analysis of database. These include the
neglect of chain connectivity in the reference state, and the
problematic implicit assumption of Boltzmann distribution
\citep{Thomas&Dill96_JMB,Thomas&Dill96_PNAS,Ben-Naim97}. We defer a
detailed discussion to Section~\ref{subsec:statistical}.

An alternative method to develop potential functions for proteins
is by optimization. For example, in protein design, we can use the
thermodynamic hypothesis of Anfinsen to require that the native
amino acid sequence $\ba_N$ mounted on the native structure
$\bs_N$ has the best (lowest) fitness score compared to a set of
alternative sequences (sequence decoys) taken from unrelated
proteins known to fold into a different fold ${\cal D} = \{\bs_N,
\ba_D\}$ when mounted on the same native protein structure
$\bs_N$:
\[
H(f(\bs_N, \ba_N)) < H(f(\bs_N, \ba_D)) \quad \mbox{for all }
(\bs_N, \ba_D) \in {\cal D}.
\]
Equivalently, the native sequence will have the highest
probability to fit into the specified native structure. This is
the same principle described in
\citep{ShakhGutin93_PNAS,Deutsch96_PRL,Li96_Science}. Sometimes we
can further require that the score difference must be greater than
a constant $b>0$ \citep{Shakhnovich94_PRL}:
\[
H(f(\bs_N, \ba_N)) + b < H(f(\bs_N, \ba_D)) \quad \mbox{for all }
(\bs_N, \ba_D) \in {\cal D}.
\]

Similarly, for protein structure prediction and protein folding,
we require that the native amino acid sequence $\ba_N$ mounted on
the native structure $\bs_N$ has the lowest energy compared to a
set of alternative conformations
 (decoys)
${\cal D} = \{\bs_D, \ba_N\}$:
\[
H(f(\bs_N, \ba_N)) < H(f(\bs_D, \ba_N)) \quad \mbox{for all }
\bs_D \in {\cal D}.
\]
and
\[
H(f(\bs_N, \ba_N)) + b < H(f(\bs_D, \ba_S)) \quad \mbox{for all }
(\bs_D, \ba_N) \in {\cal D}.
\]
when we insist to maintain an energy gap between the native
structure and decoy conformations.  For linear potential function,
we have:
\begin{equation}
\bw \cdot \bc_N + b < \bw \cdot \bc_D \quad \mbox{for all } \bc_D
= f(\bs_D, \ba_N) \label{eq:linearOptimal}
\end{equation}
Our goal is to find a set of parameters through optimization for
the potential function such that all these inequalities are
satisfied.

As discussed earlier, there are three key steps in developing
effective knowledge-based scoring function using optimization: (1) the
functional form, (2) the generation of a large set of decoys for
discrimination, and (3) the optimization techniques. The initial
step of choosing an appropriate functional form is important.
Knowledge-Based pairwise potential functions are usually all in the form
of weighted linear sum of interacting residue pairs. In this
functional form, the weight coefficients are the parameters of the
potential function, which are optimized for discrimination. This
is the same functional form used in statistical potential, where
the weight coefficients are derived from database statistics. The
objectives of optimization are often maximization of energy gap
between native protein and the average of decoys, or energy gap
between native and decoys with lowest score, or the $z$-score of
the native protein
\citep{Goldstein92_PNAS,Maiorov&Crippen92_JMB,Thomas&Dill96_PNAS,Koretke96,Hao96_PNAS,Mirny&Shakhnovich96_JMB,Vendruscolo98_JCP,Koretke98_PNAS,Tobi&Elber00_Proteins_1,Vendruscolo00_Proteins,Dima00_PS,Micheletti01_Proteins,Bastolla01_Proteins}.

\subsection{Geometric nature of discrimination}

\begin{figure}[thb]
\centerline{\epsfig{figure=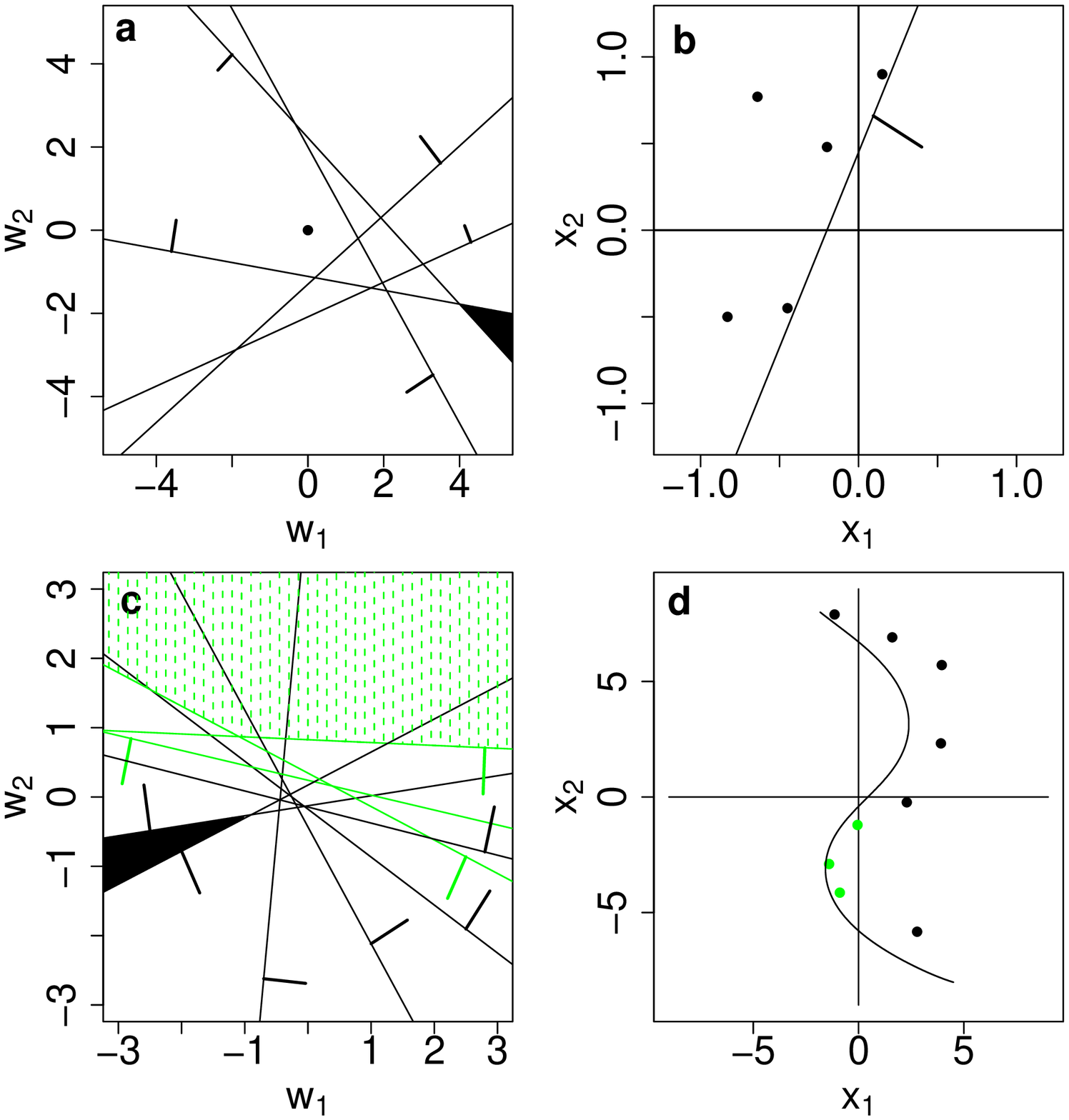,width=3.5in}} \caption[This
is what shows up in the List of Figures.]{\small \sf Geometric
views of the inequality requirement for protein scoring function.
Here we use a two-dimensional toy example for illustration. (a).
In the first geometric view, the space $\real^2$ of $\bw = (w_1,
w_2)$ is divided into two half-spaces by an inequality
requirement, represented as a hyperplane $\bw \cdot (\bc_N -
\bc_D) + b < 0$. The hyperplane, which is a line in $\real^2$, is
defined by the normal vector $(\bc_N - \bc_D)$, and its distance
$b/||\bc_N - \bc_D ||$ from the origin. In this figure, this
distance is set to 1.0. The normal vector is represented by a
short line segment whose direction points away from the straight
line. A feasible weight vector $\bw$ is located in the half-space
opposite to the direction of the normal vector $(\bc_N - \bc_D)$.
With the given set of inequalities represented by the lines, any
weight vector $\bw$ located in the shaped polygon can satisfy all
inequality requirement and provides a linear scoring function that
has perfect discrimination. (b). A second geometric view of the
inequality requirement for linear protein scoring function. The
space $\real^2$ of $\bx=(x_1, x_2)$, where $\bx \equiv (\bc_N -
\bc_D)$, is divided into two half-spaces by the hyperplane $\bw
\cdot (\bc_N - \bc_D) + b < 0$. Here the hyperplane is defined by
the normal vector $\bw$ and its distance $b/||\bw ||$ from the
origin. The origin corresponds to the native protein. All points
$\{\bc_N - \bc_D\}$ are located on one side of the hyperplane away
from the origin, therefore satisfying the inequality requirement.
That is, a linear scoring function $\bw$ such as the one
represented by the straight line in this figure can have perfect
discrimination. (c). In the second toy problem, a set of
inequalities are represented by a set of straight lines according
to the first geometric view. A subset of the inequalities require
that the weight vector $\bw$ to be located in the shaded convex
polygon on the left, but another subset of inequalities require
that $\bw$ to be located in the dashed convex polygon on the top.
Since these two polygons do not intersect, there is no weight
vector $\bw$ that can satisfy all inequality requirements. That
is, no linear scoring function can classify these decoys from
native protein. (d). According to the second geometric view, no
hyperplane can separate all points $\{\bc_N - \bc_D\}$ from the
origin. But a nonlinear curve formed by a mixture of Gaussian
kernels can have perfect separation of all vectors $\{ \bc_N -
\bc_D\}$ from the origin: It has perfect discrimination. }
\label{Fig:geom1}
\end{figure}

There is a natural geometric view of the inequality requirement
for weighted linear sum scoring functions. A useful observation is
that each of the inequalities divides the space of $\real^d$ into
two halves separated by a hyperplane (Figure~4.4a). The
hyperplane for Equation~\ref{eq:linearOptimal} is defined by the
normal vector $(\bc_N - \bc_D)$ and its distance $b/||\bc_N -
\bc_D ||$ from the origin. The weight vector $\bw$ must be located
in the half-space opposite to the direction of the normal vector
$(\bc_N - \bc_D)$. This half-space can be written as $\bw \cdot
(\bc_N - \bc_D) + b < 0$. When there are many inequalities to be
satisfied simultaneously, the intersection of the half-spaces
forms a convex polyhedron \citep{Edels87}. If the weight vector is
located in the polyhedron, all the inequalities are satisfied.
Scoring functions with such weight vector $\bw$ can discriminate
the native protein sequence from the set of all decoys. This is
illustrated in Figure~4.4a for a two-dimensional toy
example, where each straight line represents an inequality $\bw
\cdot (\bc_N - \bc_D) + b < 0$ that the scoring function must
satisfy.

For each native protein $i$, there is one convex polyhedron ${\cal
P}_i$ formed by the set of inequalities associated with its
decoys. If a scoring function can discriminate simultaneously $n$
native proteins from a union of sets of sequence decoys, the
weight vector $\bw$ must be located in a smaller convex polyhedron
$\cal P$ that is the intersection of the $n$ convex polyhedra:
\[
\bw \in {\cal P} = \bigcap_{i=1}^n {\cal P}_i.
\]

There is yet another geometric view of the same inequality
requirements. If we now regard $(\bc_N - \bc_D)$ as a point in
$\real^d$, the relationship $\bw \cdot (\bc_N - \bc_D) + b < 0$
for all sequence decoys and native proteins requires that all
points $\{\bc_N - \bc_D\}$ are located on one side of a different
hyperplane, which is defined by its normal vector $\bw$ and its
distance $b/||\bw||$ to the origin (Figure~4.4b). We can
show that such a hyperplane exists if the origin is not contained
within the convex hull of the set of points $\{ \bc_N - \bc_D\}$
(see Appendix).

The second geometric view looks very different from the first
view. However, the second view is dual and mathematically
equivalent to the first geometric view. In the first view, a point
$\bc_N - \bc_D$ determined by the structure-decoy pair $c_N =
(\bs_N,\ba_N)$ and $c_D=(\bs_N, \ba_D)$ corresponds to a
hyperplane representing an inequality, a solution weight vector
$\bw$ corresponds to a point located in the final convex
polyhedron. In the second view, each structure-decoy pair is
represented as a point $ \bc_N -\bc_D$ in $\real^d$, and the
solution weight vector $\bw$ is represented by a hyperplane
separating all the points ${\cal C} = \{\bc_N - \bc_D \}$ from the
origin.

\subsection{Optimal linear potential function} Several optimization
methods have been applied to find the weight vector $\bw$ of
linear scoring function. The Rosenblantt perceptron method works
by iteratively updating an initial weight vector $\bw_0$
\citep{Vendruscolo98_JCP,Micheletti01_Proteins}. Starting with a
random vector, {\it e.g.}, $\bw_0 ={\bf 0}$, one tests each native
protein and its decoy structure. Whenever the relationship $ \bw
\cdot (\bc_N - \bc_D) + b < 0$ is violated, one updates $\bw$ by
adding to it a scaled violating vector $\eta \cdot (\bc_N -
\bc_D)$. The final weight vector is therefore a linear combination
of protein and decoy count vectors:
\begin{equation}
\bw = \sum \eta (\bc_N - \bc_D) = \sum_{N \in {\cal N}} \alpha_N
\bc_N - \sum_{D \in {\cal D}} \alpha_D \bc_D. \label{LinearEq}
\end{equation}
Here $\cal N$ is the set of native proteins, and $\cal D$ is the
set of decoys. The set of coefficients $\{\alpha_N \} \cup
\{\alpha_D\}$ gives a dual form representation of the weight
vector $\bw$, which is an expansion of the training examples
including both native and decoy structures.

According to the first geometric view, if the final convex
polyhedron $\cal P$ is non-empty, there can be infinite number of
choices of $\bw$, all with perfect discrimination. But how do we
find a weight vector $\bw$ that is optimal? This depends on the
criterion for optimality. For example, one can choose the weight
vector $\bw$ that minimizes the variance of score gaps between
decoys and natives:
$$ \arg_\bw \min
\frac{1}{|{\cal D}|} \sum \left( \bw \cdot (c_N - c_D)\right)^2 -
\left[ \frac{1}{|{\cal D}|}\sum_D \left(\bw \cdot (\bc_N
-\bc_D)\right) \right]^2
$$
as used in reference \citep{Tobi&Elber00_Proteins_1}, or minimizing
the $Z$-score of a large set of native proteins, or minimizing the
$Z$-score of the native protein and an ensemble of decoys
\citep{Chiu&Goldstein98_FD,Mirny&Shakhnovich96_JMB}, or maximizing
the ratio $R$ between the width of the distribution of the score
and the average score difference between the native state and the
unfolded ones \citep{Goldstein92_PNAS,Hao99}. A series of important
works using perceptron learning and other optimization techniques
\citep{Friedrichs&Wolynes89_Science,Goldstein92_PNAS,Tobi&Elber00_Proteins_1,Vendruscolo98_JCP,Dima00_PS}
showed that effective linear sum scoring functions can be
obtained.

There is another optimality criterion according to the second
geometric view \citep{Hu&Liang04_Bioinformatics}. We can choose the
hyperplane $(\bw, b)$ that separates the set of points $\{\bc_N -
\bc_D\}$ with the largest distance to the origin. Intuitively, we
want to characterize proteins with a region defined by the
training set points $\{\bc_N - \bc_D \}$. It is desirable to
define this region such that a new unseen point drawn from the
same protein distribution as $\{\bc_N - \bc_D \}$ will have a high
probability to fall within the defined region. Non-protein points
following a different distribution, which is assumed to be
centered around the origin when no {\it a priori\/} information is
available, will have a high probability to fall outside the
defined region. In this case, we are more interested in modeling
the region or support of the distribution of protein data, rather
than estimating its density distribution function. For linear
scoring function, regions are half-spaces defined by hyperplanes,
and the optimal hyperplane $(\bw, b)$ is then the one with maximal
distance to the origin. This is related to the novelty detection
problem and single-class support vector machine studied in
statistical learning theory
\citep{VapChe64,VapChe74,ScholkopfSmola02}. In our case, any
non-protein points will need to be detected as outliers from the
protein distribution characterized by $\{\bc_N - \bc_D \}$. Among
all linear functions derived from the same set of native proteins
and decoys, an optimal weight vector $\bw$ is likely to have the
least amount of mis-labellings. The optimal weight vector $\bw$
can be found by solving the following quadratic programming
problem:
\begin{eqnarray}
\mbox{Minimize } & \frac{1}{2} || \bw||^2
\\
\mbox{subject to} & \bw \cdot (\bc_N - \bc_D) + b < 0 \mbox{ for
all } N \in {\cal N} \mbox{ and } D \in {\cal D}.
\label{Eqn:PrimalLinear}
\end{eqnarray}
The solution maximizes the distance $b/||\bw||$ of the plane
$(\bw, b)$ to the origin. We obtained the solution by solving the
following support vector machine problem:
\begin{equation}
\begin{array}{ll}
\mbox{Minimize} & \frac{1}{2}\|\bw\|^2 \\
\mbox{subject to} &\bw \cdot \bc_N + d \le -1 \\
&\bw \cdot \bc_D +d \ge 1,
\end{array}
\label{Eq:svm}
\end{equation}
where $d>0$. Note that a solution of Problem (\ref{Eq:svm})
satisfies the constraints in Inequalities
(\ref{Eqn:PrimalLinear}), since subtracting the second inequality
here from the first inequality in the constraint conditions of
(\ref{Eq:svm}) will give us $\bw \cdot (\bc_N - \bc_D) +2 \le 0$.

\subsection{Optimal nonlinear potential function}
\label{sec:nonlinear} Optimal linear potential function can be
obtained using the optimization strategy discussed above.
However, it is possible that the weight vector $\bw$ does not
exist, {\it i.e.}, the final convex polyhedron ${\cal P} =
\bigcap_{i=1}^n {\cal P}_i$ may be an empty set.  This occurs if a
large number of native protein structures are to be simultaneously
stabilized against a large number of decoy conformations, no such
potential functions in the linear functional form can be found
\citep{Vendruscolo00_Proteins,Tobi&Elber00_Proteins_1}.

According to our geometric pictures, there are two possible
scenarios. First, for a specific native protein $i$, there may be
severe restriction from some inequality constraints, which makes
${\cal P}_i$ an empty set. Some decoys are very difficult to
discriminate due to perhaps deficiency in protein representation.
In these cases, it is impossible to adjust the weight vector so
the native protein has a lower score than the sequence decoy.
Figure~4.4c shows a set of inequalities represented by
straight lines according to the first geometric view. In this
case, there is no weight vector that can satisfy all these
inequality requirements. That is, no linear scoring function can
classify all decoys from native protein. According to the second
geometric view (Figure~4.4d), no hyperplane can
separate all points (black and green) $\{\bc_N - \bc_D\}$ from the
origin, which corresponds to the native structures.

Second, even if a weight vector $\bw$ can be found for each native
protein, {\it i.e.}, $\bw$ is contained in a nonempty polyhedron,
it is still possible that the intersection of $n$ polyhedra is an
empty set, {\it i.e.}, no weight vector can be found that can
discriminate all native proteins against the decoys
simultaneously. Computationally, the question whether a solution
weight vector $\bw$ exists can be answered unambiguously in
polynomial time \citep{Karmarkar84}. If a large number ({\it e.g.},
hundreds) of native protein structures are to be simultaneously
stabilized against a large number of decoy conformations ({\it
e.g.}, tens of millions), no such potential functions can be found
computationally
\citep{Vendruscolo00_Proteins,Tobi&Elber00_Proteins_1}. Similar
conclusion is drawn in a study for protein design, where it was
found that no linear potential function can simultaneously
discriminate a large number of native proteins from sequence
decoys~\citep{Hu&Liang04_Bioinformatics}.

A fundamental reason for such failure is that the functional form
of linear sum is too simplistic. It has been suggested that
additional descriptors of protein structures such as higher order
interactions ({\it e.g.}, three-body or four-body contacts) should
be incorporated in protein description
\citep{Betancourt99_PS,Munson97,Zheng97}. Functions with polynomial
terms using up to 6 degree of Chebyshev expansion has also been
used to represent pairwise interactions in protein folding
\citep{Fain02_PS}.

We now discuss an alternative approach. Let us still limit
ourselves to pairwise contact interactions, although it can be
naturally extended to include three or four body interactions
\citep{Li&Liang05_Proteins}. We can introduce a nonlinear scoring
function analogous to the dual form of the linear function in
Equation (\ref{LinearEq}), which takes the following form:
\begin{equation}
H(f(\bs, \ba)) = H(\bc) = \sum_{D \in {\cal D}}\alpha_D K(\bc,
\bc_D) - \sum_{N \in {\cal N}}\alpha_N K(\bc, \bc_N),
\label{nonlinear}
\end{equation}
where $\alpha_D \ge 0$ and $\alpha_N \ge 0$ are parameters of the
scoring function to be determined, and $\bc_D = f(\bs_N, \ba_D)$
from the set of decoys $\cal D = \{ (\bs_N, \ba_D )\}$ is the
contact vector of a sequence decoy $D$ mounted on a native protein
structure $\bs_N$, and $\bc_N = f(\bs_N, \ba_N)$ from the set of
native training proteins ${\cal N} = \{(\bs_N, \ba_N) \}$ is the
contact vector of a native sequence $\ba_N$ mounted on its native
structure $\bs_N$. In this study, all decoy sequence $\{ \ba_D\}$
are taken from real proteins possessing different fold structures.
The difference of this functional form from linear function in
Equation (\ref{LinearEq}) is that a kernel function $K(\bx, \by)$
replaces the linear term. A convenient kernel function $K$ is:
\[K(\bx, \by) =
e^{-||\bx -\by||^2/2\sigma^2} \quad \mbox{for any vectors $\bx$
and $\by \in {\cal N} \bigcup {\cal D}$},
\]
where $\sigma^2$ is a constant. Intuitively, the surface of the
scoring function has smooth Gaussian hills of height $\alpha_D$
centered on the location $\bc_D$ of decoy protein $D$, and has
smooth Gaussian cones of depth $\alpha_N$ centered on the location
$\bc_N$ of native structures $N$. Ideally, the value of the
scoring function will be $-1$ for contact vectors $\bc_N$ of
native proteins, and will be $+1$ for contact vectors $\bc_D$ of
decoys. \label{subsec:nonlinear}

\subsection{Deriving optimal nonlinear scoring function.} To obtain the
nonlinear scoring function, our goal is to find a set of
parameters $\{\alpha_D, \alpha_N\}$ such that $ H(f(\bs_N,
\ba_N))$ has value close to $-1$ for native proteins, and the
decoys have values close to $+1$. There are many different choices
of $\{\alpha_D, \alpha_N \}$. We use an optimality criterion
originally developed in statistical learning theory
\citep{Vapnik95,Burges98,ScholkopfSmola02}. First, we note that we
have implicitly mapped each structure and decoy from $\real^{210}$
through the kernel function of $K(\bx, \by) = e^{-||\bx
-\by||^2/2\sigma^2}$ to another space with dimension as high as
tens of millions. Second, we then find the hyperplane of the
largest margin distance separating proteins and decoys in the
space transformed by the nonlinear kernel. That is, we search for
a hyperplane with equal and maximal distance to the closest native
proteins and the closest decoys in the transformed high
dimensional space. Such a hyperplane can be found by obtaining the
parameters $\{ \alpha_D \}$ and $\{ \alpha_N \}$ from solving the
following Lagrange dual form of quadratic programming problem:
\begin{eqnarray*}
\mbox{Maximize } & \sum_{ \substack{ i\in {\cal N} \cup {\cal D},
} } \alpha_i - \frac{1}{2}\sum_{ \substack{ i,j\in {{\cal N}\cup
{\cal D}} } } y_i y_j \alpha_i \alpha_j e^{-||c_i -
\bc_j||^2/2\sigma^2}
\\
\mbox{subject to} & 0\le \alpha_i \le C,\\
\end{eqnarray*}
where $C$ is a regularizing constant that limits the influence of
each misclassified protein or decoy
\citep{VapChe64,VapChe74,Vapnik95,Burges98,ScholkopfSmola02}, and
$y_i =-1$ if $i$ is a native protein, and $y_i= +1$ if $i$ is a
decoy. These parameters lead to optimal discrimination of an
unseen test set
\citep{VapChe64,VapChe74,Vapnik95,Burges98,ScholkopfSmola02}. When
projected back to the space of $\real^{210}$, this hyperplane
becomes a nonlinear surface. For the toy problem of
Figure~4.4, Figure~4.4d shows that such a
hyperplane becomes a nonlinear curve in $\real^2$ formed by a
mixture of Gaussian kernels. It separates perfectly all vectors
$\{ \bc_N - \bc_D\}$ (black and green) from the origin. That is, a
nonlinear scoring function can have perfect discrimination.

\subsection{Optimization techniques.}
The techniques that have been used for optimizing potential
function include perceptron learning, linear programming, gradient
descent, statistical analysis, and support vector machine
\citep{Tobi&Elber00_Proteins_1,Vendruscolo00_Proteins,Xia&Levitt00_JCP,Bastolla00_PNAS,Bastolla01_Proteins,Hu&Liang04_Bioinformatics}.
These are standard techniques that can be found in optimization
and machine learning literature.  For example, there are excellent
linear programming solvers based on simplex method, as implemented
in {\sc Clp}, {\sc Glpk}, and {\sc lp\_solve}~\citep{LP}, and based
on interior point method as implemented in the {\sc Bpmd}
\citep{Meszaros96_CMA}, the {\sc Hopdm} and the {\sc PCx}
packages~\citep{PCx}. We neglect the details of these techniques
and point readers to the excellent treatises of
\citep{Papadimitriou98,Vanderbei96}.

\section{Applications}
Knowledge-Based potential function has been widely used in the study of
protein structure prediction, protein folding, and protein-protein
interaction. In this section, we discuss briefly some of these
applications.  Additional details of applications of knowledge-based
potential can be found in other chapters of this book.

\begin{figure}[!t]
\centerline{\epsfig{figure=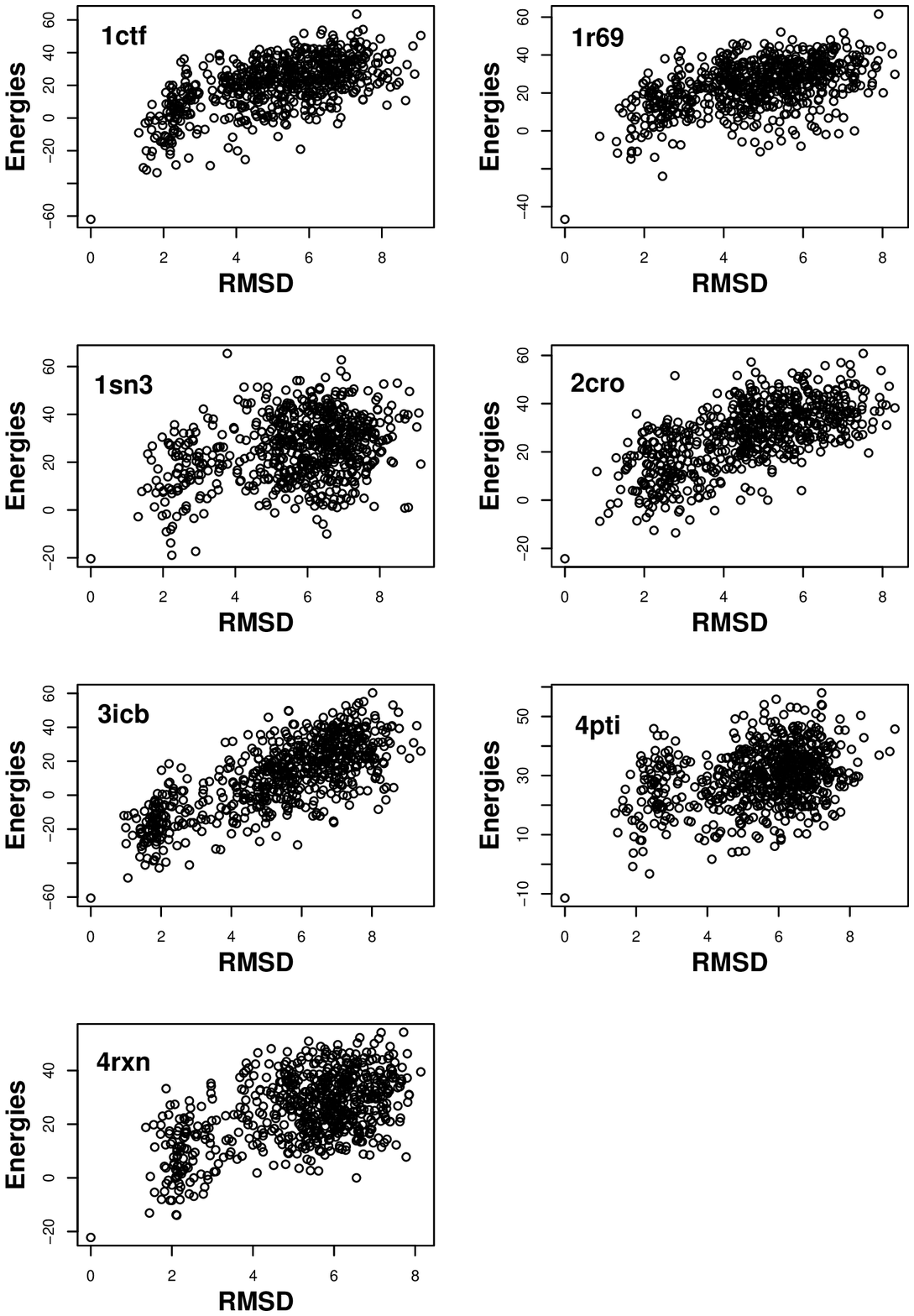,width=4.5in}} \caption{\small
\sf Energies evaluated by packing and distance dependent residue
contact potential plotted against the RMSD to native structures
for conformations in Park\& Levitt Decoy Set. } \label{Fig:4state}
\end{figure}

\subsection{Protein structure prediction}
Protein structure prediction is an extraordinarily complex task
that involves two major components: sampling the conformational
space and recognizing the near native structures from the ensemble
of sampled conformations.

In protein structure prediction, methods for conformational
sampling will generate a huge number of candidate protein
structures.  These are often called {\it decoys}.  Among these
decoys, only a few are near native structures that are very
similar to the native structure.  An knowledge-based potential function
must be used to discriminate the near native structures from all
other decoys for a successful structure prediction.

Several decoy sets have been developed
which are used as objective benchmarks to test if an knowledge-based
potential function can successfully identify the native and near
native structures.  For example, Park and Levitt (1996)
constructed a 4-state-reduced decoy set. This decoy test set
contains native and near-native conformations of seven sequences,
along with about 650 misfolded structures for each sequence.  The
positions of $C_\alpha$ of these decoys were generated by
exhaustively enumerating ten selectively chosen residues in each
protein using a 4-state off-lattice model. All other residues were
assigned the phi/psi value based on the best fit of a 4-state
model to the native chain~\citep{ParkLevitt96_JMB}.

A central depository of folding decoy conformations is the {\sc Decoys
R'Us} \citep{DecoyRus}.  See Section~\ref{Sec:URL} for the {\sc url}
links to download several folding and docking decoy sets.  A variety
of knowledge-based potential functions have been developed and their
performance in decoy discrimination have steadily improved
~\citep{Zhou02_ProSci,Lu&Skolnick01_Proteins,Li&Liang03_Proteins}.

Figure~4.5 shows an example of decoy discrimination
on the {\it 4-state-reduced} decoy set. This result is based on
the residue-level packing and distance-dependent alpha potential
discussed earlier. For all of the seven proteins in the
4-state-reduced set, the native structures have the lowest energy.
In addition, all of the decoys with the lowest energy are within
2.5 $\mathring{\mbox{A}}$ RMSD to the native structure.

Table~\ref{Tab:Performance} lists the performance of the geometric
potential function in folding and docking decoy discriminations.
Several studies examine the comparative performance of different
knowledge-based potential functions
\citep{ParkLevitt96_JMB,Zhou02_ProSci,Gilis04_JBSD}.  Such
evaluations often are based on measuring the success in ranking native
structure from a large set of decoy conformations and in obtaining a
large $z$-score) for the native protein structure.  Because the
development of potential function is a very active research field, the
comparison of performances of different potential functions will be
different as new models and techniques are developed and incorporated.

\begin{table}[!t]
\setlength{\tabcolsep}{1pt}
\renewcommand{\arraystretch}{1.5}
\begin{scriptsize}
\begin{center}
\caption{\small Performance of geometric potential on folding and
docking decoy discrimination.}\label{Tab:Performance}
\begin{tabular}{l c c c c c c c c c c c c c c}
\hline
\multirow{3}{0.42in}{Folding\\decoy\\sets}&\multicolumn{2}{c}{4-state-reduced}
&&\multicolumn{2}{c}{lattice-ssfit}  &&\multicolumn{2}{c}{fisa-casp3}&&\multicolumn{2}{c}{fisa} &&\multicolumn{2}{c}{lmds}\\
\cline{2-3} \cline{5-6} \cline{8-9} \cline{11-12} \cline{14-15}
& Native$^a$ &$z^b$ && Native &$z$ && Native &$z$ && Native &$z$ && Native &$z$\\
&  7/7  &  4.46 &&8/8 & 7.70 &&3/3 & 5.23&&3/4 & 5.42 && 7/10 & 1.45\\
\hline
\multirow{4}{0.42in}{Docking\\decoy\\sets}&\multicolumn{2}{b{0.8in}}{Rosetta-Bound-Perturb} &&\multicolumn{2}{b{0.9in}}{Rosetta-Unbound-Perturb}&&\multicolumn{2}{b{0.8in}}{Rosetta-Unbound-Global}&&\multicolumn{2}{c}{Vakser's}&&\multicolumn{2}{c}{Sternberg's}\\
\cline{2-3} \cline{5-6} \cline{8-9} \cline{11-12} \cline{14-15}
& Native &$z$ && Native &$z$ && Native &$z$ && Native &$z$ && Native &$z$\\
&  50/54 & 12.75 &&53/54 & 12.88 && 53/54 & 8.55 && 4/5  &  4.45 &&16/16 & 4.45\\
\cline{2-15} &\multicolumn{2}{l}{RDOCK} & \multicolumn{12}{l}{ 29/42$^c$}\\
\hline

\multicolumn{15}{p{4.5in}}{\footnotesize $^a$ Number of native
structures ranking first. {\it eg.} $7/7$ means seven out of seven
native structures have the lowest energy among their corresponding
decoy sets. }\\
\multicolumn{15}{p{4.5in}}{\footnotesize $^b$:  $z = \overline{E}
- E_{native}/{\sigma}$; $\overline{E}$ and $\sigma$ are the mean
and standard deviation of the energy values of conformations,
respectively.} \\
\multicolumn{15}{p{4.5in}}{\footnotesize $^c$:  Native complex is
not included in this docking decoy sets. 32 out of 42 decoy sets
have at least one near native structures (cRMSD $< 2.5
 \mathring{\mbox{ A}}$) in the top 10 structures.}\\

\end{tabular}
\end{center}
\end{scriptsize}
\end {table}

Knowledge-based potential function can not only be applied at the end
of the conformation sampling  to recognize near native structures, it can also
be used  during conformation generation to guide
the efficient sampling of protein structures.  Details of this
application can be found in~\citep{Jernigan96_COSB,Scheraga99_COSB}.
In addition, knowledge-based potential also plays important role in
protein threading studies.   Chapter 13 provides further detailed
discussion.

\subsection{Protein-protein docking prediction}
Knowledge-Based potential function can also be used to study
protein-protein interactions.  Here we give an example of
predicting the binding surface of seven antibody or antibody
related proteins ({\it e.g.}, Fab fragment, T-cell receptor)
\citep{Li&Liang05_PSB}.  These protein-protein complexes are taken
from the 21 {\sc Capri} (Critical Assessment of PRedicted
Interactions) target proteins.  {\sc Capri} is a community-wide
competition designed to objectively assess the abilities in
protein-protein docking prediction \citep{CAPRI}. In {\sc Capri}, a
blind docking prediction starts from two known crystallographic or
NMR structures of unbound proteins and ends with a comparison to a
solved structure of the protein complex, to which the participants
did not have access.  Knowledge-Based potential functions, together with
geometric complementarity scoring functions, can be used to
recognize near native docking complexes and to guide the
generation of conformations for protein-protein docking.

\begin{figure}[thb]
\begin{center}
$\begin{array}{ccc} \epsfig{figure=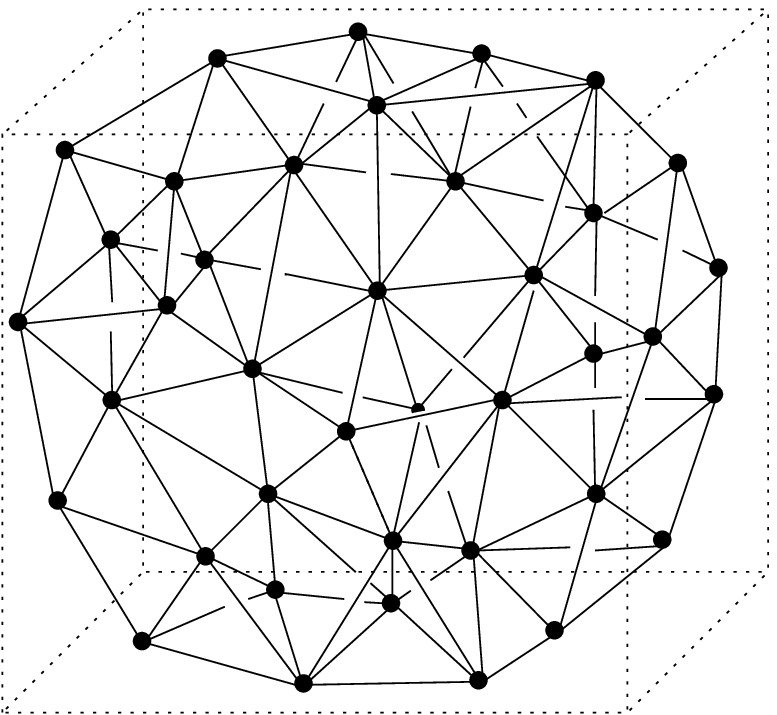,height=1.4in} &
\epsfig{figure=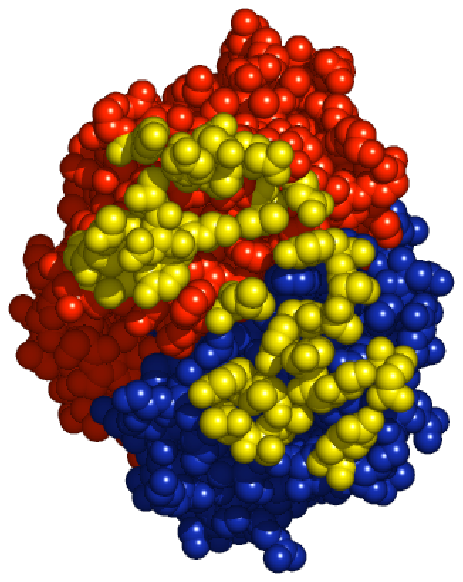,height=1.4in}&
\epsfig{figure=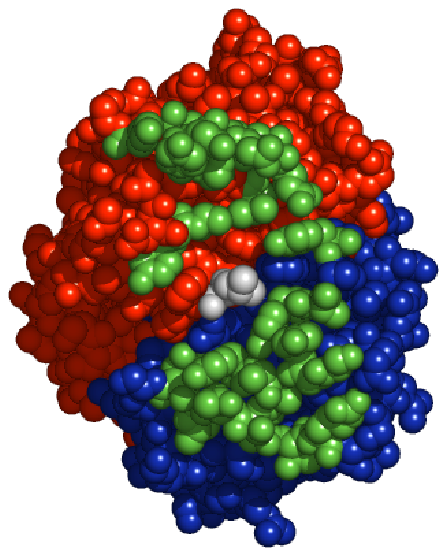,height=1.4in} \\
&\mbox{Native antibody interface} & \mbox{Best scored patch}\\[0.2in]
{\mbox{\bf (a)} } & \multicolumn{2}{c}{\mbox{\bf (b)} }
\end{array}$
\label{Fig:Fig2} \caption{\small \sf Recognition of binding
surface patch of {\sc Capri} targets.  (a) Boundary of alpha shape
for a {\it cargo} protein. Each node represents a surface residue,
and each edge represents the alpha edge between two surface
residues. A candidate surface patch is generated by starting from
a surface residue on the cargo protein, and following alpha edges
on the boundary of the alpha shape by breadth-first search, until
$m$ residues are included.  (b) Native interface and the surface
patch with the best score on the antibody of the protein complex
CAPRI Target T02. Only heavy chain (in red) and light chain (in
blue) of the antibody are drawn.  The antigen is omitted from this
illustration for clarity.  The best scored surface patch (in
green) resembles the native interface (in yellow): 71\% residues
from this surface patch are indeed on the native binding
interface. The residue in white is the starting residue used to
generate this surface patch with the best score.}
\end{center}
\end{figure}

When docking two proteins together, we say a {\it cargo} protein
is docked to a fixed {\it seat} protein.  To determine the binding
surfaces on the cargo protein, we can examine all possible surface
patches on the unbound structure of cargo protein as candidate
binding interfaces. The alpha knowledge-based potential function is then
used to identify native or near native binding surfaces.  To
evaluate the performance of the potential function, we assume the
knowledge of the binding interface on the seat protein.  We
further assume the knowledge of the degree of near neighbors for
interface residues.

We first partition the surface of the unbound cargo protein into
candidate surface patches, each has the same size as the native
binding surface of $m$ residues.  A candidate surface patch is
generated by starting from a surface residue on the cargo protein,
and following alpha edges on the boundary of the alpha shape by
breadth-first search, until $m$ residues are found
(Figure~4.6a).  We construct $n$ candidate surface
patches by starting in turn from each of the $n$ surface residue
on the cargo protein.  Because each surface residue is the center
of one of the $n$ candidate surface patch, the set of candidate
surface patches cover exhaustively the whole protein binding
interface.

\begin{table}[!t]
\setlength{\tabcolsep}{4.5pt}
\begin{scriptsize}
\begin{center}
\caption{\small Recognition of native binding surface of CAPRI
targets.} \label{Tab:Table2} \vspace*{3mm}
\begin{tabular}{l l r  c c r c}
\hline
&&\multicolumn{2}{c}{ Antibody$^a$ }&&\multicolumn{2}{c}{Antigen }\\
\cline{3-4}\cline{6-7}
Target& Complex & Rank$_{native}^b$ & Overlap$^c$  && Rank$_{native}$ & Overlap\\
\hline
T02   & Rotavirus VP6-Fab    &   1/283$^d$  &   0.71     &&1/639   &0.68\\
T03   & Flu hemagglutinin-Fab&   1/297  &   0.56     &&1/834   &0.71\\
T04   & $\alpha$-amylase-camelid Ab VH 1 &56/89 &0.60&&102/261 &0.03\\
T05   & $\alpha$-amylase-camelid Ab VH 2 &23/90 &0.57&&57/263  &0.25\\
T06   & $\alpha$-amylase-camelid Ab VH 3 &1/88  &0.70&&1/263   &0.62\\
T07   & SpeA superantigen TCR$\beta$     &1/172 &0.57&&1/143   &0.61\\
T13   & SAG1-antibody complex&1/286     &   0.64 &&1/249       &0.69\\
\hline \\
\multicolumn{7}{l}{\footnotesize $^a$ ``Antibody'': Different
surface patches on the antibody molecule
are evaluated by the}\\
\multicolumn{7}{l}{\footnotesize scoring function, while the
native binding surface on the antigen remains unchanged.
}\vspace*{-0pt}\\
\multicolumn{7}{l}{\footnotesize ``Antigen'':
similarly defined as ``Antibody''.}\vspace*{2pt}\\
\multicolumn{7}{l}{\footnotesize $^b$ Ranking of the
native binding surface among all candidate surface patches.}\vspace*{2pt}\\
\multicolumn{7}{l}{\footnotesize $^c$ Fraction of residues from
the best candidate surface patch that overlap with residues}\\
\multicolumn{7}{l}{\footnotesize from the
native binding surface patch.} \vspace*{2pt}\\
\multicolumn{7}{l}{\footnotesize $^d$ The first number is the rank
of native binding surface and the second number is the
}\vspace*{-0pt}\\ \multicolumn{7}{l}{\footnotesize number of total
candidate surface patches.}\\ \hline\\
\end{tabular}
\end{center}
\vspace*{-5mm}
\end{scriptsize}
\end {table}

Second, we assume that a candidate surface patch on the cargo
protein has the same set of contacts as that of the native binding
surface. The degree of near neighbors for each hypothetical
contacting residue pair is also assumed to be the same.  We
replace the $m$ residues of the native surface with the $m$
residues from the candidate surface patch. There are
$\frac{m!}{\prod^{20}_{i=1}m_i!}$ different ways to permute the
$m$ residues of the candidate surface patch, where $m_i$ is the
number of residue type $i$ on the candidate surface patch.  A
typical candidate surface patch has about 20 residues, therefore
the number of possible permutation is very large. For each
candidate surface patch, we take a sample of 5,000 random
permutations.  For a candidate surface patch $SP_i$, we assume
that the residues can be organized so that they can interact with
the binding partner at the lowest energy. Therefore, the binding
energy $E(SP_i)$ is estimated as: $$E(SP_i) = \min_k E{(SP_i)}_k,
\quad k = 1, \cdots, 5,000.$$ Here $E{(SP_i)}_k$ is calculated
based on the residue-level packing and distance-dependent
potential for the $k$-th permutation. The value of $E(SP_i)$ is
used to rank the candidate surface patches.

We assess the statistical potential by taking antibody/antigen
protein in turn as the seat protein, and the antigen/antibody as
cargo protein.  The native interface on the seat protein is fixed.
We test if our statistical potential can discriminate native surface
patch on the cargo protein from the set of candidate surface
patches.  We also test if the best scored patch resembles the
native patch.
The results are listed in Table~\ref{Tab:Table2} and the predicted
antigen-binding interface of target T02 is shown in
Figure~4.6(b) as an example. For five out of the seven
protein complexes, we succeeded in discriminating the native
patches on both the antibody and the antigen.  Over 50\% of the
residues from the best scored patch overlaps with corresponding
native patch.  Our statistical potential does not work as well for
the target T04 and T05, because the antibodies of these two
complexes do not use their CDR domains to recognize the antigens
as an antibody usually does, and such examples are not present in
the dataset of the 34 antibody-antigen complexes, based on which
the alpha potential function was obtained.

\subsection{Protein design}
Protein design aims to identify sequences compatible with a given
protein fold but incompatible to any alternative folds
\citep{KoehlLevitt99_JMB,KoehlLevitt99b_JMB}. The goal is to design a
novel protein that may not exist in nature but has enhanced or novel
biological function.  Several novel proteins have been successfully
designed in recent years
\citep{DahiyatMayo97_Science,HillDeGrado00_ACR,Looger03_Nature,Baker03_Science}.
The problem of protein design is complex, because even a small protein
of just 50 residues can have an astronomical number of sequences
($10^{65}$) This clearly precludes exhaustive search of the sequence
space with any computational or experimental method. Instead, protein
design methods rely on potential functions for biasing the search
towards the feasible regions that encode protein sequences.  To select
the correct sequences and to guide the search process, a design
potential function is critically important.  Such a scoring function
should be able to characterize the global fitness landscape of many
proteins simultaneously.

Here, we briefly describe the application of optimal nonlinear
design potential function discussed in Section~\ref{sec:nonlinear}
\citep{Hu&Liang04_Bioinformatics} in protein design.  We aim to
solve a simplified protein sequence design problem. Our goal is to
distinguish each native sequence for a major portion of
representative protein structures from a large number of
alternative decoy sequences, each a fragment from proteins of
different fold.

To train the nonlinear potential function, a list of 440 proteins
was compiled from the 1998 release ({\sc Whatif98}) of the {\sc
Whatif} database \citep{Vendruscolo00_Proteins}.  Using gapless
threading \citep{Maiorov&Crippen92_JMB}, a set of 14,080,766
sequence decoys was obtained.  The entries in {\sc Whatif99}
database that are not present in {\sc Whatif98} are used as a test
set.  After clean-up, the test set consists of 194 proteins and
3,096,019 sequence decoys.

To test the design scoring functions for discriminating native
proteins from sequence decoys, we take the sequence $\ba$ from the
conformation-sequence pair $(\bs_N, \ba)$ for a protein with the
lowest score as the predicted sequence.  If it is not the native
sequence $\ba_N$, the discrimination failed and the design scoring
function does not work for this protein.

The nonlinear design scoring function is capable of discriminating
all of the 440 native sequences. In contrasts, no linear scoring
function can succeed in this task. The nonlinear potential
function also works well for the test set, where it succeeded in
correctly identifying 93.3\% (181 out of 194) of native sequences
in the independent test set of 194 proteins.  This compares
favorably with results obtained using optimal linear folding
scoring function taken as reported in
\citep{Tobi&Elber00_Proteins_1}, which succeeded in identifying
80.9\% (157 out of 194) of this test set.  It also has better
performance than optimal linear scoring function based on
calculations using parameters reported in reference
\citep{Bastolla01_Proteins}, which succeeded in identifying 73.7\%
(143 out of 194) of proteins in the test set.  The
Miyazawa-Jernigan statistical potential succeeded in identifying
113 native proteins out of 194) (success rate 58.2\%).

\subsection{Protein stability and binding affinity}
Because the stability of protein in the native conformation is
determined by the distribution of the full ensemble of
conformations, namely, the partition function $Z(\ba)$ of the
protein sequence $\ba$, care must be taken when using statistical
potentials to compare the stabilities of different protein
sequences adopting the same given conformation as in protein
design~\citep{Miyazawa&Jernigan96_JMB,Sippl90_JMB}.  This issue is
discussed in some detail in Subsection~\ref{Subsec:Boltz_Sum}.

Nevertheless, it is expected that statistical potential should work
well in estimating protein stability changes upon mutations, as the
change in partition functions of the protein sequence is small.
In most such studies and studies using physics-based empirical
potential (see  Chapter 4 in this book and reference
\citep{Bordner&Abagyan04_Proteins}), good correlation coefficient
(0.6--0.8) between predicted and measured stability change can be
achieved~\citep{Gilis&Rooman96_JMB,Gilis&Rooman97_JMB,Guerois&Serrano02_JMB,Bordner&Abagyan04_Proteins,Hoppe&Schomburg05_ProSci,Zhou02_ProSci}.

Several studies have shown that statistical potentials can also be
used to predict quantitative binding free energy of
protein-protein or protein-ligand
interactions~\citep{Shakhnovich96_JACS,Thornton99_JCC,Muegge&Martin99_JMC,Zhou04_Proteins,Zhang&Zhou05_JMC}.
In fact, Xu {\it et al.\/} showed that a simple number count of
hydrophilic bridges across the binding interface is strongly
correlated with binding free energies of protein-protein
interaction~\citep{Xu&Nussinov97_JMB}. This study suggests that
binding free energy may be predicted successfully by number counts
of various types of interfacial contacts defined using some
distance threshold.  Such number count studies provide an excellent
benchmark to quantify the improvement in predicting binding free
energy when using statistical potentials for different
protein-protein and protein-ligand complexes.  Similar to
prediction of protein stability change upon mutation, knowledge
based potential function played an important role in a successful
study of predicting binding free energy changes upon mutation
\citep{Kortemme&Baker02_PNAS,Kortemme&Baker04_SS}.

\section{Online resource}
A list of online sources of decoy data for folding and docking
is provided in Table~\ref{Tab:URL}. \label{Sec:URL}
\begin{table}[!t]
\setlength{\tabcolsep}{1.2pt}
\renewcommand{\arraystretch}{1.2}
\begin{small}
\begin{center}
\caption{\small Database of folding and docking decoy sets.}
\label{Tab:URL} \vspace*{3mm}
\begin{tabular}{c  c  c >{\tt}p{2.4in}}
\hline
Decoy sets & Type && \multicolumn{1}{c}{URL} \\
\hline
Decoy `R' Us$^a$ & folding && http://dd.stanford.edu/\\
Loop         & folding && http://francisco.compbio.ucsf.edu /$^\sim$jacobson/decoy.htm\\
CASP         & folding && http://predictioncenter.org/\\
ZDOCK, RDOCK  & docking && http://zlab.bu.edu/$^\sim$leely/RDOCK \_decoy/\\
Vakser decoy set     & docking && http://www.bioinformatics.ku.edu/ files/vakser/decoys/\\
Sternberg decoy set  & docking && http://www.sbg.bio.ic.ac.uk/docking/ all\_decoys.html\\
Rosetta      & docking, folding && http://depts.washington.edu/bakerpg/\\
CAPRI        & docking && http://capri.ebi.ac.uk/\\
\hline

\multicolumn{4}{p{4.2in}}{$^a$: The database of Decoy `R' Us contains
{\it multiple decoy sets}, {\it Single decoy sets} and {\it loop decoy
sets}. {\it 4-state-reduced} decoy set is included in the multiple
decoy sets.

}
\end{tabular}
\end{center}
\end{small}
\end{table}

\section{Discussion}
\subsection{Knowledge-Based statistical potential functions}
\label{subsec:statistical} The statistical potential functions are
often derived based on several assumptions: (a) protein energetics
can be decomposed primarily into pairwise interactions; (b)
interactions are independent from each other; (c) the partition
function in native proteins $Z$ and in reference states $Z'$ are
approximately equal; (d) the probability of occupancy of a state
follows the Boltzmann distribution. These assumptions are often
unrealistic and raise questions about the validity of the
statistical potential functions: Can statistical potential
functions provide energy-like quantities such as the folding free
energy of a protein, or the binding free energy of a
protein-protein complex~\citep{Thomas&Dill96_JMB}?\,  Can statistical
potential functions correctly recognize the native structures from
alternative conformations?

\vspace*{.15 in} \noindent {\bf The assumptions of statistical
knowledge-based potential functions.} From Equation~\ref{eq:21}, we can
obtain the potential function $H(\bc)$ by estimating the
probability $\pi(\bc)$. However, we need a number of assumptions
for this approach to work. We need the independency assumption to
have:
$$
\pi(\bc) = \prod_i \pi(c_i) = \prod_i \prod_{c_i} \pi_i,
$$
where $c_i$ is the number of occurrence of $i$-th structural
feature, {\it e.g.}, number of a specific residue pair contact;
$\pi_i$ is the probability of $i$-th structural feature in the
database. That is, we have to assume that the distribution of a
specific structural feature is independent and not influenced by
any other features, and is of no consequence for the distribution
of other features as well. We also need to assume that $\bc$
provides an adequate characterization of protein interactions, and
the functional form of $\bw \cdot \bc$ provides the correct
measurement of the energy of the interactions. We further need to
assume that the energy for a protein-solvent system is
decomposable, {\it, i.e.}, the overall energy can be partitioned
into many basic energy terms, such as pairwise interactions,
desolvation energies. Moreover, the partition functions $Z'$ in a
chosen reference state are approximately equal to the partition
functions $Z$ in native proteins. These above assumptions together
lead to the Boltzmann assumption that the structural features
contained in the protein database must be a population correctly
sampled under the Boltzmann distribution. That is, for for any
protein descriptor, we have:
$$
\pi_i \propto \exp (-w_i).
$$

To calculate $\pi_i$ in practice, we have to rely on another
assumption that all protein structures are crystallized at the same
temperature. Therefore, the distribution $\pi_i$ is reasonably
similar for all proteins in the database, and hence the frequency
counts of protein descriptors in different protein structures can
be combined by simple summation with equal weight.

Clearly, none of these assumptions are strictly true. However, the
success of many applications of using the statistical knowledge-based
potentials indicate that they do capture many important properties
of proteins. The question for improving statistical potential
function is, how seriously each of these assumptions is violated
and to what extent it affects the validity of the potential
function. A few assumptions specific to a particular potential
function (such as the coordination and solvation assumptions for
the Miyazawa-Jernigan's reaction model) have been described
earlier. Here we discuss several assumptions in details below.

\vspace*{.15 in} \noindent {\bf Interactions are not independent.}
Using a HP (hydrophobic-Polar) model on two-dimensional lattice,
Thomas and Dill (1996) tested the accuracy of Miyazawa-Jernigan
contact potentials and Sippl's distance-dependent potentials. In
HP model, a peptide chain contains only two types of monomer: $H$
and $P$. The true energies are set as $H_{(H,H)} = -1, H_{(H,P)} =
0$ and $H_{(P,P)} = 0$. Monomers are in contact if they are
non-bonded nearest neighbors on the lattice. The conformational
space was exhaustively searched for all sequences with the chain
length from 11 to 18. A sequence is considered to have a native
structure if it has a unique ground energy state. All native
structures were collected to build a structure database, from
which the statistical potentials are extracted by following the
Miyazawa-Jernigan or the Sippl method. The extracted energies are
denoted as $e_{(H,H)}, e_{(H,P)}$, and $e_{(P,P)}$.

It was found that neither of these two methods can extract the
correct energies. All extracted energies by these two methods
depend on chain length, while the true energies do not. Using
Miyazawa-Jernigan's method, the $(H,H)$ contact is correctly
determined as dominant and attractive. However, the estimated
values for $e_{(H,P)}$ and $e_{(P,P)}$ are not equal to zero,
whereas the true energies $H_{(H,P)}$ and $H_{(P,P)}$ are equal to
zero. Using Sippl's method, the extracted potentials erroneously
show a distance-dependence, {\it i.e.}, $(H,H)$ interactions are
favorable in short-distance but unfavorable in long-distance, and
conversely for $(P,P)$ interactions, whereas the true energies in
the HP model only exist between a first-neighbor $(H,H)$ contact,
and become zero for all the interactions separated by two or more
lattice units.

These systematic errors result from the assumption that the
pairwise interactions are independent, and thus the volume
exclusion in proteins can be neglected~\citep{Thomas&Dill96_JMB}.
However, $(H,H)$ interactions indirectly affects the observed
frequencies of $(H,P)$ and $(P,P)$ interactions. First, in both
contact and distance-dependent potentials, because only a limited
number of inter-residue contacts can be made within the restricted
volume at a given distance, the high density of $(H,H)$ pairs at
short distances is necessarily coupled with the low density
(relative to reference state) of $(H,P)$ and $(P,P)$ pairs at the
same distances, especially at the distance of one lattice unit. As
a result, the extracted $(H,P)$ and $(P,P)$ energies are
erroneously unfavorable at short distance. Second, for
distance-dependent potentials, the energy of a specific type of
pair interaction at a given distance is influenced by the same
type of pair at different distances. For example, the high density
of $(H,H)$ pairs at short distances causes a compensating
depletion (relative to the uniform density reference state) at
certain longer distances, and conversely for $(H,P)$ and $(P,P)$
interactions. Admittedly this study was carried out using models
of short chain lengths and a simple alphabet of residues where the
foldable sequences may be very homologous, hence the observed
artifacts are profound, the deficiencies of the statistical
potentials revealed in this study such as the excluded volume
effect is likely to be significant in potential functions derived
from real proteins.

\vspace*{.15 in} \noindent {\bf Pairwise interactions are not
additive.} Interactions stabilizing proteins are often modeled by
pairwise contacts at atom or residue level. An assumption
associated with this approach is the additivity of pairwise
interactions, namely, the total energy or fitness score of a
protein is the linear sum of all of its pairwise interactions.

However, the non-additivity effects have been clearly demonstrated
in cluster formation of hydrophobic methane molecules both in
experiment~\citep{Ben-Naim97_JCP} and in
simulation~\citep{Rank&Baker97_ProSci,Shimizu&Chan01_JCP,Shimizu&Chan02_Proteins,Czaplewski&Scheraga00_ProtSci}.
Protein structure refinement will likely require higher order
interactions \citep{Betancourt99_PS}. Some three-body contacts have
been introduced in several studies
\citep{Eastwood&Wolynes01_JCP,Rossi02_BJ,Godzik92_PNAS,Godzik92_JMB},
where physical models explicitly incorporating three-body
interactions are developed. In addition, several studies of
Delaunay four-body interactions clearly showed the importance of
including higher order interactions in explaining the observed
frequency distribution of residue contacts
\citep{Tropsha03_Bioinformatics,Tropsha01_JMB,Tropsha01_Proteins,Zheng&Tropsha97_PSB,Singh&Tropsha96_JCB,Munson97}.

Li and Liang (2005) introduced a geometric model based on the
Delaunay triangulation and alpha shape to collect three-body
interactions in native proteins. A nonadditivity coefficient
$\nu_{(i,j,k)}$ is introduced to compare the three-body potential
energy $e_{(i,j,k)}$ with the summation of three pairwise
interactions $e_{i,j}, e_{(i,k)},$ and $e_{(j,k)}$:
$$
\nu_{(i,j,k)} = \exp[-e_{(i,j,k)}]/\exp[ -(e_{(i,j)} + e_{(i,k)} +
e_{(j,k)})].
$$

There are three possibilities: (1) $\nu = 1$: interaction of a
triplet type is additive in nature and can be well approximated by
the sum of three pairwise interactions; (2) $\nu > 1$: three-body
interactions are cooperative and their association is more
favorable than three independent pairwise interactions; (3) $\nu <
1$: three-body interactions are anti-cooperative.

After systematically quantifying the nonadditive effects of all
$1,540$ three-body contacts, it was found that hydrophobic
interactions and hydrogen bonding interactions make nonadditive
contributions to protein stability, but the nonadditive nature
depends on whether such interactions are located in protein
interior or on protein surface. When located in interior, many
hydrophobic interactions such as those involving alkyl residues
are anti-cooperative, namely $\nu < 1$. Salt-bridge and regular
hydrogen-bonding interactions such as those involving ionizable
residues and polar residues are cooperative in interior. When
located on protein surface, these salt-bridge and regular hydrogen
bonding interactions are anti-cooperative with $\nu<1$, and
hydrophobic interactions involving alkyl residues become
cooperative~\citep{Li&Liang05_Proteins}.

\vspace*{.25in} \noindent {\bf Sequence dependency of the
partition function {\boldmath $Z(\ba)$}.}\label{Subsec:Boltz_Sum} We can obtain the total
effective energy $\Delta E(\bs,\ba)$ given a structure
conformation $\bs$ and its amino acid sequence $\ba$ from
Equation~(\ref{eq:Z}):
\begin{equation}
\begin{split}
\Delta H(f(\bs,\ba)) &= \Delta H(\bc) = \sum_i \Delta H(c_i)
\\
&= -kT \sum_{c_i} \ln (\frac{\pi(c_i)}{\pi'(c_i)}) -kT \ln
(\frac{Z(\ba)}{Z'(\ba)})
\end{split}
\end{equation}
where $c_i$ is the total number count of the occurrence of the
$i$-th descriptor, {\it e.g.}, the total number of $i$-th type of
pairwise contact. The summation involving $Z(\ba)$ and $Z'(\ba)$
is ignored during the evaluation of $\Delta H(c_i)$ by assuming
$Z(\ba) \approx Z'(\ba)$.

It is clear that both $Z(\ba)$ and $Z'(\ba)$ do not depend on the
particular structural conformation $\bs$. Therefore, the omission
of the term of the partition functions $ -kT \ln
(\frac{Z(\ba)}{Z'(\ba)})$ will not affect the rank ordering of
energy values of different conformations ({\it i.e.}, decoys) for
the same protein sequence.  On the other hand, it is also clear
that both $Z(\ba)$ and $Z'(\ba)$ depend on the specific sequence
$\ba$ of a protein. Therefore, there is no sound theoretical basis
to compare the stabilities between different proteins using the
same knowledge-based potential function, unless the ratio of
$Z(\ba)/Z'(\ba)$ for each individual sequence is known and is
included during the evaluation
\citep{Miyazawa&Jernigan85_M,Samudrala&Moult98_JMB,Sippl90_JMB}.
Notably, {\sc Dfire} and other statistical energy functions have been
successful used to predict binding affinities across different
protein-protein/peptide complexes. Nevertheless, the theoretical
basis is not sound either, because the values of partition
function Z(a)s for different protein complexes can be drastically
different. It remains to be seen whether a similarly successful
prediction of binding affinities can be achieved just by using the
number of native interface contacts at some specific distance
interval, i.e. the packing density along the native interface.
This omission is probably benign for the problem of predicting
free energy change of a protein monomer or binding free energy
change of a protein-protein complex upon point mutations, because
the distribution of the ensemble of protein conformations may not
change significantly after one or several point mutations.

\vspace*{.25in} \noindent {\bf Evaluating potential function.}
  The measure used for
performance evaluation of potential functions is important.  For
example, $z$-score of native protein among decoys is widely-used as an
important performance statistic.  However, $z$-score strongly depends
on the properties of the decoy set.  Imagine we have access to the
true energy function. If a decoy set has a diverse distribution in
true energy values, the $z$-score of the native structure will not be
very large.  However, this by no means suggests that a knowledge-based
energy function that gives a larger $z$-score for native protein is
better than the true energy function.  Alternative measures may
provide more accurate or useful performance evaluation.  For example,
the correlation $r$ of energy value and {\sc crmsd} may be helpful in
protein structure prediction.  Since a researcher has no access to the
native structure, (s)he has to rely on the guidance of an energy
function to search for better structures with lower {\sc crmsd} to the
unknown native structure.  For this purpose, a potential function with
a large $r$ will be very useful.  Perhaps the performance of a
potential function should be judged not by a single statistic but
comprehensively by a number of measures.

\subsection{Relationship of knowledge-based energy functions and further development}
The Miyazawa-Jernigan contact potential is the first widely used
knowledge-based potential function. Because it is limited by the
simple spatial description of a cut-off distance, it cannot capture
the finer spatial details.  Several distance-dependent potentials have
been developed to overcome this limitation, and in general have better
performance
\citep{Lu&Skolnick01_Proteins,Samudrala&Moult98_JMB,Zhou02_ProSci}.  A
major focus of works in this area is the development of models for the
reference state.  For example, the use of the ideal gas as reference
state in the potential function {\sc Dfire} significantly improves the
performance in folding and docking decoy
discrimination~\citep{Zhang&Zhou04_ProSci}.

Because protein surface, interior, and protein-protein interface are
packed differently, the propensity of the same pairwise interaction
can be different depending on whether the residues are solvent-exposed
or are buried.  The contact potential of Simons {\it et al.\/}
considers two types of environment, {\it i.e.}, buried and non-buried
environments separately \citep{Simons&Baker99_Proteins}.  The geometric
potential function \citep{Li&Liang05_PSB} described in
Subsection~\ref{subsec:geom} incorporates both dependencies on
distance and fine-graded local packing, resulting in significant
improvement in performance.  Table~\ref{Tab:Performance} show that
this potential can be successfully used in both protein structure and
docking prediction.
Knowledge based potential has also been developed to account for the loss of
backbone,  side-chain, and translational entropies in folding and binding
\citep{Amzel00_ME,Amzel94_Proteins}.

Another emphasis of  recent development of potential function is
the orientational dependency of pairwise
interaction~\citep{Baker03_JMB,Thirumalai03_JCP,Thirumalai04_ProSci,Miyazawa&Jernigan05_JCP}.
Kortemme {\it et al.\/} developed an orientation-dependent hydrogen
bonding potential, which improved prediction of protein structure and
specific protein-protein interactions~\citep{Baker03_JMB}.  Miyazawa and
Jernigan developed a fully anisotropic distance-dependent potential,
with drastic improvements in decoy discrimination over the original
Miyazawa-Jernigan contact potential~\citep{Miyazawa&Jernigan05_JCP}.

\vspace*{.25in}\noindent {\bf Computational Efficiency}
Given current computing power, all potential functions discussed above
can be applied to large-scale discrimination of native or near-native
structures from decoys. For example, the geometric potential requires
complex computation of the Delaunay tetrahedrization and alpha shape
of the molecule (see Chapter 7 for details).  Nevertheless, the time
complexity is only ${\cal O}(N\log N)$, where $N$ is the number of
residues for residual-level potentials or atoms for atom-level
potentials.  For comparison, a naive implementation of
contact computing without the use of proper data structure such as a quad-tree
or $k$-d tree is $ {\cal O} (N^2)$.

In general, atom-level potentials have better accuracy in recognizing
native structures than residue-level potentials, and is often
preferred for the final refinement of predicted structures, but it is
computationally too expensive to be applicable in every step of a
folding or sampling computation.

\vspace*{.25in} \noindent {\bf Potential function for membrane protein.}
 The potential
functions we have discussed in Section 3 are based on the structures of
soluble proteins. Membrane proteins are located in a very different
physico-chemical environment.  They also have different amino acid
composition, and they fold differently.  Potential functions developed
for soluble proteins are therefore not applicable to membrane
proteins. For example, Cys-Cys has the strongest pairing propensity
because of the formation of disulfide bond.  However, Cys-Cys pairs
rarely occur in membrane proteins.  This and other difference in
pairwise contact propensity between membrane and soluble proteins are
discussed in \citep{Adamian01_JMB}.

Nevertheless, the physical models underlying most potential functions
developed for soluble proteins can be modified for membrane
proteins~\citep{Adamian01_JMB,Adamian02_Prot,Adamian03_JMB,Park_Proteins04,Jackups05_JMB}.
For example, Sale {\it et al\/} used the {\sc Mhip} potential
developed in \citep{Adamian01_JMB} to predict optimal bundling of TM
helices.  With the help of 27 additional sparse distance constraints
from experiments reported in literature, these authors succeeded in
predicting the structure of dark-adapted rhodopsin to within 3.2 \AA\
of the crystal structure \citep{Sale_PS04}.  It is likely that
statistical potentials can be similarly developed for protein-ligand
and protein-nucleotides interactions using the same principle.

\subsection{Optimized potential function}
Knowledge based potential function derived by optimization
has a number of characteristics that are distinct from statistical
potential.  We discuss in detail below.

\vspace*{.15 in} \noindent {\bf Training set for optimized
potential function. } Unlike statistical potential functions where
each native protein in the database contribute to the knowledge-based
scoring function, only a subset of native proteins contribute. In
an optimized potential function, in addition, a small fraction of
decoys also contribute to the scoring function.  In the study
of~\citep{Hu&Liang04_Bioinformatics}, about $ 50\% $ of native
proteins and $<0.1\%$ of decoys from the original training data of
440 native proteins and 14 million sequence decoys contribute to
the potential function.

As illustrated in the second geometric views, the discrimination
of native proteins occurs at the boundary surface between the
vector points and the origin. It does not help if the majority of
the training data are vector points away from the boundary
surface. This implies the need for optimized potential to have
appropriate training data. If no {\it a priori\/} information is
known, it is likely many decoys ($>$millions) will be needed to
accurately define the discrimination boundary surface, because of
the usually large dimension of the descriptors for proteins.
However, this imposes significant computational burden.

Various strategies have been developed to select only the most
relevant vector points.  Usually, one may only include the most
difficult decoys during training, such as decoys with lower energy
than native structures, decoys with lowest absolute energies, and
decoys already contributing to the potential function in previous
iteration
\citep{Micheletti01_Proteins,Tobi&Elber00_Proteins,Hu&Liang04_Bioinformatics}.
In addition, an iterative training process is often necessary
\citep{Micheletti01_Proteins,Tobi&Elber00_Proteins,Hu&Liang04_Bioinformatics}.

\vspace*{.15 in} \noindent {\bf Reduced nonlinear potential
function.} The use of nonlinear terms for potential function
involves large datasets, because they are necessary {\it a
priori\/} to define accurately the discrimination surface.  This
demands the solution of a huge optimization problem.  Moreover,
the representation of the boundary surface using a large basis set
requires expensive computing time for the evaluation of a new
unseen contact vector $\bc$.  To overcome these difficulties,
non-linear potential function needs to be further simplified.

One simple approach is to use alternative optimal criterion, for
example, by minimizing the distance expressed in 1-norm instead of
the standard 2-norm Euclidean distance.  The resulting potential
function will automatically have reduced terms. Another promising
approach is to use rectangle kernels (Hu, Dai, and Liang,
manuscript).

\vspace*{.15 in} \noindent {\bf Potential function by optimal
regression.}  Currently, most optimized potential functions are
derived based on decoy discrimination, which is a form of binary
classification.  Here we suggest a conceptual improvement that can
significantly improve the development of optimized potential
functions. If we can measure the thermodynamic stabilities of all
major representative proteins under identical experimental
conditions ({\it e.g.}, temperature, pH, salt concentration, and
osmolarity), we can attempt to develop potential functions with
the objective of minimizing the regression errors of fitted energy
values and measured energy values.  The resulting energy surface
will then provide quantitative information about protein
stabilities. However, the success of this strategy will depend on
coordinated experimental efforts in protein thermodynamic
measurements. The scale of such efforts may need to be similar to
that of genome sequencing projects and structural genomics
projects.

\subsection {Data dependency of knowledge-based potentials}
There are many directions to improve knowledge-based potential functions.
Often it is desirable to include additional descriptors in the energy
functions to more accurately account for solvation, hydrogen bonding,
backbone conformation ({\it e.g.}, $\phi$ and $\psi$ angles), and
side chain entropies. Furthermore, potential functions with different
descriptors and details may be needed for different tasks ({\it e.g.}
backbone prediction {\it vs} structure refinement,
\citep{Rohl&Baker04_ME}).

An important issue in both statistical potential and optimized
potential is their dependency on the amount of available training data
and possible bias in such data.  For example, whether a
knowledge-based potential derived from a bias data set is applicable
to a different class of proteins is the topic of several
studies~\citep{Zhou04_BiophysJ,Khatun04_JMB}.  Zhang {\it et al\/}
further studies the effect of database choice on statistical potential
\citep{Zhang_BJ04}. In addition, when the amount of data is limited,
over-fitting is a real problem if too many descriptors are introduced
in either of the two types of potential functions.  For statistical
potential, hierarchical hypothesis testing should help to decide whether
additional terms is warranted.  For optimized potential,
cross-validation will help to uncover possible overfitting
\citep{Hu&Liang04_Bioinformatics}.

\section{Summary}
In this chapter, we discussed the general framework of developing
knowledge-based potential functions in terms of molecular descriptors,
functional form, and parameter calculations. We also discussed the
 underlying thermodynamic hypothesis of protein folding.  With the
assumption that frequently observed protein features in a database
of structures correspond to low energy state, frequency of
observed interactions can be converted to energy terms.  We then
described in details the models behind the Miyazawa-Jernigan
contact potential, distance dependent potentials, and geometric
potentials.  We also discussed how to weight sample structures of
varying degree of sequence similarity in the structural database.
In the section of optimization method, we describe general
geometric models for the problem of obtaining optimized knowledge-based
potential functions, as well as methods for developing optimal
linear and nonlinear potential functions.  This is followed by a
brief discussion of several applications of the knowledge-based
potential functions. Finally, we point out general limitations and
possible improvements for the statistical and optimized potential
functions.

\section{Further reading}
Anfinsen's thermodynamic hypothesis can be found in
\citep{Anfinsen61_PNAS,Anfinsen73_Science}. More technical details
of the Miyazawa-Jernigan contact potential are described in
\citep{Miyazawa&Jernigan85_M,Miyazawa&Jernigan96_JMB}.  Distance
dependent potential function was first proposed by Sippl in
\citep{Sippl90_JMB}, with further development described in
\citep{Lu&Skolnick01_Proteins,Samudrala&Moult98_JMB}.  The
development of geometric potentials can be found in
\citep{Zheng&Tropsha97_PSB,Tropsha01_JMB,Li&Liang03_Proteins,Tropsha03_Bioinformatics,McConkey03_PNAS}.
The gas-phase approximation of the reference state is discussed in
\citep{Zhou02_ProSci}.  Thomas and Dill offered insightful comments
about the deficiency of knowledge-based statistical potential functions
\citep{Thomas&Dill96_JMB}.  The development of optimized linear
potential functions can be found in
\citep{Vendruscolo00_Proteins.pair,Micheletti01_Proteins,Tobi&Elber00_Proteins_1}.
The geometric view for designing optimized potential function and
the nonlinear potential function are based on the results in
\citep{Hu&Liang04_Bioinformatics}.

\section{Acknowledgments}
We thank Drs.\ Bob Jernigan, Hui Lu, Dong Xu, Hongyi
Zhou, and Yaoqi Zhou for helpful discussions. This work is supported
by grants from the National Science Foundation (CAREER DBI0133856),
the National Institute of Health (GM68958), the Office of Naval
Research (N000140310329), and the Whitaker Foundation (TF-04-0023).


\begin{thebibliography}{139}
\expandafter\ifx\csname natexlab\endcsname\relax\def\natexlab#1{#1}\fi
\expandafter\ifx\csname url\endcsname\relax
  \def\url#1{{\tt #1}}\fi
\expandafter\ifx\csname urlprefix\endcsname\relax\def\urlprefix{URL }\fi

\bibitem[{Adamian {\em et~al.\/}(2003)Adamian, Jackups, Binkowski, and
  Liang}]{Adamian03_JMB}
Adamian L, Jackups R, Binkowski TA, Liang J (2003) {Higher-order interhelical
  spatial interactions in membrane proteins}. J Mol Biol 327:251--272.

\bibitem[{Adamian and Liang(2001)}]{Adamian01_JMB}
Adamian L, Liang J (2001) {Helix-helix packing and interfacial pairwise
  interactions of residues in membrane prot eins}. J Mol Biol 311:891--907.

\bibitem[{Adamian and Liang(2002)}]{Adamian02_Prot}
Adamian L, Liang J (2002) {Interhelical hydrogen bonds and spatial motifs in
  membrane proteins: polar clamps and serine zippers}. Proteins 47:209--218.

\bibitem[{Amzel(2000)}]{Amzel00_ME}
Amzel LM (2000) Calculation of entropy changes in biological processes:
  folding, binding, and oligomerization. Methods Enzymol 323:167--77.

\bibitem[{Anfinsen {\em et~al.\/}(1961)Anfinsen, Haber, Sela, and
  White}]{Anfinsen61_PNAS}
Anfinsen C, Haber E, Sela M, White F (1961) The kinetics of formation of native
  ribonuclease during oxidation of the reduced polypeptide chain. Proc Natl
  Acad Sci 47:1309--1314.

\bibitem[{Anfinsen(1973)}]{Anfinsen73_Science}
Anfinsen CB (1973) Principles that govern the folding of protein chains.
  Science 181:223--230.

\bibitem[{Bastolla {\em et~al.\/}(2001)Bastolla, Farwer, Knapp, and
  Vendruscolo}]{Bastolla01_Proteins}
Bastolla U, Farwer J, Knapp EW, Vendruscolo M (2001) How to guarantee optimal
  stability for most representative structurs in the protein data bank.
  Proteins 44:79--96.

\bibitem[{Bastolla {\em et~al.\/}(2000)Bastolla, Vendruscolo, and
  Knapp}]{Bastolla00_PNAS}
Bastolla U, Vendruscolo M, Knapp EW (2000) A statistical mechanical method to
  optimize energy functions for protein folding. Proc Natl Acad Sci USA
  97:3977--3981.

\bibitem[{Ben-Naim(1997{\natexlab{a}})}]{Ben-Naim97_JCP}
Ben-Naim A (1997{\natexlab{a}}) Statistical potentials extracted from protein
  structures: Are these meaningful potentials? J Chem Phys 107:3698--3706.

\bibitem[{Ben-Naim(1997{\natexlab{b}})}]{Ben-Naim97}
Ben-Naim A (1997{\natexlab{b}}) Statistical potentials extracted from protein
  structures: {A}re these meaningful potentials? J Chem Phys 107:3698--3706.

\bibitem[{Berkelaar(2004)}]{LP}
Berkelaar M (2004) {LP\_Solve package} .
\newline\urlprefix\url{http://www.cs.sunysb.edu/~algorith/implement/lpsolve/
  implement.shtml}

\bibitem[{Betancourt and Thirumalai(1999)}]{Betancourt99_PS}
Betancourt MR, Thirumalai D (1999) Pair potentials for protein folding:
  {C}hoice of reference states and sensitivity of predicted native states to
  variations in the interaction schemes. Protein Sci 8:361--369.

\bibitem[{Bienkowska {\em et~al.\/}(1999)Bienkowska, Rogers, and
  Smith}]{Smith99_Proteins}
Bienkowska JR, Rogers RG, Smith TF (1999) Filtered neighbors threading.
  Proteins 37:346--359.

\bibitem[{Bordner and Abagyan(2004)}]{Bordner&Abagyan04_Proteins}
Bordner AJ, Abagyan RA (2004) Large-scale prediction of protein geometry and
  stability changes for arbitrary single point mutations. Proteins
  57(2):400--13.

\bibitem[{Buchete {\em et~al.\/}(2003)Buchete, Straub, and
  Thirumalai}]{Thirumalai03_JCP}
Buchete NV, Straub JE, Thirumalai D (2003) Anisotropic coarse-grained
  statistical potentials improve the ability to identify nativelike protein
  structures. The Journal of Chemical Physics 118:7658--7671.

\bibitem[{Buchete {\em et~al.\/}(2004)Buchete, Straub, and
  Thirumalai}]{Thirumalai04_ProSci}
Buchete NV, Straub JE, Thirumalai D (2004) Orientational potentials extracted
  from protein structures improve native fold recognition. Protein Sci
  13:862--74.

\bibitem[{Burges(1998)}]{Burges98}
Burges CJC (1998) A {T}utorial on {S}upport {V}ector {M}achines for {P}attern
  {R}ecognition. Knowledge Discovery and Data Mining 2.
\newline\urlprefix\url{/papers/Burges98.ps.gz}

\bibitem[{Carter~Jr. {\em et~al.\/}(2001)Carter~Jr., LeFebvre, Cammer, Tropsha,
  and Edgell}]{Tropsha01_JMB}
Carter~Jr. C, LeFebvre B, Cammer S, Tropsha A, Edgell M (2001) Four-body
  potentials reveal protein-specific correlations to stability changes caused
  by hydrophobic core mutations. J Mol Biol 311(4):625--638.

\bibitem[{Chan and Dill(1990)}]{HP}
Chan HS, Dill KA (1990) Origins of structure in globular proteins. Proc Natl
  Acad Sci 87:6388--6392.

\bibitem[{Chiu and Goldstein(1998)}]{Chiu&Goldstein98_FD}
Chiu TL, Goldstein RA (1998) Optimizing energy potentials for success in
  protein tertiary structure prediction. Folding Des 3:223--228.

\bibitem[{Czaplewski {\em et~al.\/}(2000)Czaplewski, Rodziewicz-Motowidlo,
  Liwo, Ripoll, Wawak, and Scheraga}]{Czaplewski&Scheraga00_ProtSci}
Czaplewski C, Rodziewicz-Motowidlo S, Liwo A, Ripoll DR, Wawak RJ, Scheraga HA
  (2000) Molecular simulation study of cooperativity in hydrophobic
  association. Protein Sci 9:1235--1245.

\bibitem[{Czyzyk {\em et~al.\/}(2004)Czyzyk, Mehrotra, Wagner, and
  Wright}]{PCx}
Czyzyk J, Mehrotra S, Wagner M, Wright S (2004) {PCx package} .
\newline\urlprefix\url{http://www-fp.mcs.anl.gov/otc/Tools/PCx/}

\bibitem[{Dahiyat and Mayo(1997)}]{DahiyatMayo97_Science}
Dahiyat BI, Mayo SL (1997) {\it De Novo\/} protein design: Fully automated
  sequence selection. Science 278:82--87.

\bibitem[{Deutsch and Kurosky(1996)}]{Deutsch96_PRL}
Deutsch JM, Kurosky T (1996) New algorithm for protein design. Phys Rev Lett
  76:323--326.

\bibitem[{DeWitte and Shakhnovich(1996)}]{Shakhnovich96_JACS}
DeWitte RS, Shakhnovich EI (1996) {SMoG: de novo design method based on simple,
  fast and accurate free energy estimates. 1. Methodology and supporting
  evidence}. J Am Chem Soc 118:11733--44.

\bibitem[{Dima {\em et~al.\/}(2000)Dima, Banavar, and Maritan}]{Dima00_PS}
Dima RI, Banavar JR, Maritan A (2000) Scoring functions in protein folding and
  design. Protein Sci 9:812--819.

\bibitem[{Dobbs {\em et~al.\/}(2002)Dobbs, Orlandini, Bonaccini, and
  Seno}]{Dobbs02_Proteins}
Dobbs H, Orlandini E, Bonaccini R, Seno F (2002) Optimal potentials for
  predicting inter-helical packing in transmembrane proteins. Proteins
  49(3):342--349.

\bibitem[{Duan and Kollman(1998)}]{Duan&Kollman98_Science}
Duan Y, Kollman PA (1998) Pathways to a protein folding intermediate observed
  in a 1-microsecond simulation in aqueous solution. Science 282:740--744.

\bibitem[{Eastwood and Wolynes(2001)}]{Eastwood&Wolynes01_JCP}
Eastwood MP, Wolynes PG (2001) Role of explicitly cooperative interactions in
  protein folding funnels: A simulation study. J Chem Phys 114(10):4702--4716.

\bibitem[{Edelsbrunner(1987)}]{Edels87}
Edelsbrunner H (1987) Algorithms in combinatorial geometry. Springer-Verlag,
  Berlin.

\bibitem[{Fain {\em et~al.\/}(2002)Fain, Xia, and Levitt}]{Fain02_PS}
Fain B, Xia Y, Levitt M (2002) Design of an optimal {C}hebyshev-expanded
  discrimination function for globular proteins. Protein Sci 11:2010--2021.

\bibitem[{Finkelstein {\em et~al.\/}(1995)Finkelstein, Badretdinov, and
  Gutin}]{whyBoltzmann}
Finkelstein AV, Badretdinov AY, Gutin AM (1995) Why do protein architectures
  have boltzmann-like statistics? Proteins 23(2):142--50.

\bibitem[{Friedrichs and Wolynes(1989)}]{Friedrichs&Wolynes89_Science}
Friedrichs MS, Wolynes PG (1989) Toward protein tertiary structure recognition
  by means of associative memory hamiltonians. Science 246:371--373.

\bibitem[{Gan {\em et~al.\/}(2001)Gan, Tropsha, and
  Schlick}]{Tropsha01_Proteins}
Gan H, Tropsha A, Schlick T (2001) Lattice protein folding with two and
  four-body statistical potentials. Proteins 43(2):161--174.

\bibitem[{Gilis(2004)}]{Gilis04_JBSD}
Gilis D (2004) Protein decoy sets for evaluating energy functions. J Biomol
  Struct Dyn 21:725--36.

\bibitem[{Gilis and Rooman(1996)}]{Gilis&Rooman96_JMB}
Gilis D, Rooman M (1996) Stability changes upon mutation of solvent-accessible
  residues in proteins evaluated by database-derived potentials. J Mol Biol
  257(5):1112--26.

\bibitem[{Gilis and Rooman(1997)}]{Gilis&Rooman97_JMB}
Gilis D, Rooman M (1997) Predicting protein stability changes upon mutation
  using database-derived potentials: solvent accessibility determines the
  importance of local versus non-local interactions along the sequence. J Mol
  Biol 272(2):276--90.

\bibitem[{Godzik {\em et~al.\/}(1992)Godzik, Kolinski, and
  Skolnick}]{Godzik92_JMB}
Godzik A, Kolinski A, Skolnick J (1992) Topology fingerprint approach to the
  inverse protein folding problem. J Mol Biol 227(1):227--238.

\bibitem[{Godzik and Skolnick(1992)}]{Godzik92_PNAS}
Godzik A, Skolnick J (1992) Sequence-structure matching in globular proteins:
  application to supersecondary and tertiary structure determination. Proc Natl
  Acad Sci 89(24):12098--102.

\bibitem[{Goldstein {\em et~al.\/}(1992)Goldstein, Luthey-Schulten, and
  Wolynes}]{Goldstein92_PNAS}
Goldstein R, Luthey-Schulten ZA, Wolynes PG (1992) Protein tertiary structure
  recognition using optimized hamiltonians with local interactions. Proc Natl
  Acad Sci USA 89:9029--9033.

\bibitem[{Guerois {\em et~al.\/}(2002)Guerois, Nielsen, and
  Serrano}]{Guerois&Serrano02_JMB}
Guerois R, Nielsen JE, Serrano L (2002) Predicting changes in the stability of
  proteins and protein complexes: a study of more than 1000 mutations. J Mol
  Biol 320(2):369--87.

\bibitem[{Hao and Scheraga(1999{\natexlab{a}})}]{Hao99}
Hao MH, Scheraga H (1999{\natexlab{a}}) Designing potential energy functions
  for protein folding. Curr Opinion Structural Biology 9:184--188.

\bibitem[{Hao and Scheraga(1996)}]{Hao96_PNAS}
Hao MH, Scheraga HA (1996) How optimization of potential functions affects
  protein folding. Proc Natl Acad Sci 93(10):4984--89.

\bibitem[{Hao and Scheraga(1999{\natexlab{b}})}]{Scheraga99_COSB}
Hao MH, Scheraga HA (1999{\natexlab{b}}) Designing potential energy functions
  for protein folding. Curr Opin Struct Biol 9:184--188.

\bibitem[{Hill {\em et~al.\/}(2000)Hill, Raleigh, Lombardi, and
  DeGrado}]{HillDeGrado00_ACR}
Hill RB, Raleigh DP, Lombardi A, DeGrado WF (2000) {\it De novo\/} design of
  helical bundles as models for understanding protein folding and function. Acc
  Chem Res 33:745--754.

\bibitem[{Hoppe and Schomburg(2005)}]{Hoppe&Schomburg05_ProSci}
Hoppe C, Schomburg D (2005) Prediction of protein thermostability with a
  direction- and distance-dependent knowledge-based potential. Protein Sci
  14:2682--92.

\bibitem[{Hu {\em et~al.\/}(2004)Hu, Li, and Liang}]{Hu&Liang04_Bioinformatics}
Hu C, Li X, Liang J (2004) Developing optimal non-linear scoring function for
  protein design. Bioinformatics 20(17):3080--98.

\bibitem[{Jackups~Jr and Liang(2005)}]{Jackups05_JMB}
Jackups~Jr R, Liang J (2005) {Interstrand Pairing Patterns in $\beta$-Barrel
  Membrane Proteins: The Positive-outside Rule, Ar omatic Rescue, and Strand
  Registration Prediction}. J Mol Biol {\it In Press}.

\bibitem[{Janicke(1987)}]{ThermoHypothesisReview}
Janicke R (1987) Folding and association of proteins. Prog Biophys Mol Biol
  49:117--237.

\bibitem[{Jernigan and Bahar(1996)}]{Jernigan96_COSB}
Jernigan RL, Bahar I (1996) Structure-derived potentials and protein
  simulations. Curr Opin Struct Biol 6:195--209.

\bibitem[{Karmarkar(1984)}]{Karmarkar84}
Karmarkar N (1984) A new polynomial-time algorithm for linear programming.
  Combinatorica 4:373--395.

\bibitem[{Karplus and Petsko(1990)}]{Karplus90_Nature}
Karplus M, Petsko GA (1990) Molecular dynamics simulations in biology. Nature
  :631--639.

\bibitem[{Khatun {\em et~al.\/}(2004)Khatun, Khare, and
  Dokholyan}]{Khatun04_JMB}
Khatun J, Khare SD, Dokholyan NV (2004) Can contact potentials reliably predict
  stability of proteins? J Mol Biol 336:1223--1238.

\bibitem[{Kocher {\em et~al.\/}(1994)Kocher, Rooman, and
  Wodak}]{Kocher&Wodak94_JMB}
Kocher JA, Rooman MJ, Wodak SJ (1994) Factors influencing the ability of
  knowledge-based potentials to identify native sequence-structure matches. J
  Mol Biol 235:1598--1613.

\bibitem[{Koehl and Levitt(1999{\natexlab{a}})}]{KoehlLevitt99_JMB}
Koehl P, Levitt M (1999{\natexlab{a}}) {\it De Novo\/} protein design. {I}.
  {I}n search of stability and specificity. J Mol Biol 293:1161--1181.

\bibitem[{Koehl and Levitt(1999{\natexlab{b}})}]{KoehlLevitt99b_JMB}
Koehl P, Levitt M (1999{\natexlab{b}}) {\it De Novo\/} protein design. {II}.
  {P}lasticity of protein sequence. J Mol Biol 293:1183--1193.

\bibitem[{Koretke {\em et~al.\/}(1996)Koretke, Luthey-Schulten, and
  Wolynes}]{Koretke96}
Koretke KK, Luthey-Schulten Z, Wolynes PG (1996) Self-consistently optimized
  statistical mechanical energy functions for sequence structure alignment.
  Protein Sci 5:1043--1059.

\bibitem[{Koretke {\em et~al.\/}(1998)Koretke, Luthey-Schulten, and
  Wolynes}]{Koretke98_PNAS}
Koretke KK, Luthey-Schulten Z, Wolynes PG (1998) Self-consistently optimized
  energy functions for protein structure prediction by molecular dynamics. Proc
  Natl Acad Sci 95(6):2932--7.

\bibitem[{Kortemme and Baker(2002)}]{Kortemme&Baker02_PNAS}
Kortemme T, Baker D (2002) A simple physical model for binding energy hot spots
  in protein-protein complexes. Proc Natl Acad Sci 99:14116--21.

\bibitem[{Kortemme {\em et~al.\/}(2004)Kortemme, Kim, and
  Baker}]{Kortemme&Baker04_SS}
Kortemme T, Kim DE, Baker D (2004) Computational alanine scanning of
  protein-protein interfaces. Sci STKE 2004:pl2.

\bibitem[{Kortemme {\em et~al.\/}(2003)Kortemme, Morozov, and
  Baker}]{Baker03_JMB}
Kortemme T, Morozov AV, Baker D (2003) An orientation-dependent hydrogen
  bonding potential improves prediction of specificity and structure for
  proteins and protein-protein complexes. J Mol Biol 326:1239--59.

\bibitem[{Krishnamoorthy and Tropsha(2003)}]{Tropsha03_Bioinformatics}
Krishnamoorthy B, Tropsha A (2003) Development of a four-body statistical
  pseudo-potential to discriminate native from non-native protein
  conformations. Bioinformatics 19(12):1540--8.

\bibitem[{Kuhlman {\em et~al.\/}(2003)Kuhlman, Dantas, Ireton, Varani,
  Stoddard, and Baker}]{Baker03_Science}
Kuhlman B, Dantas G, Ireton GC, Varani G, Stoddard BL, Baker D (2003) Design of
  a novel globular protein fold with atomic-level accuracy. Science :1364--8.

\bibitem[{Lazaridis and Karplus(2000)}]{Lazaridis&Karplus00_COSB}
Lazaridis T, Karplus M (2000) Effective energy functions for protein structure
  prediction. Curr Opin Struct Biol 10:139--145.

\bibitem[{Lee {\em et~al.\/}(1994)Lee, Xie, Freire, and
  Amzel}]{Amzel94_Proteins}
Lee KH, Xie D, Freire E, Amzel LM (1994) Estimation of changes in side chain
  configurational entropy in binding and folding: general methods and
  application to helix formation. Proteins 20:68--84.

\bibitem[{Lemer {\em et~al.\/}(1995)Lemer, Rooman, and
  Wodak}]{Lerner95_Proteins}
Lemer CMR, Rooman MJ, Wodak SJ (1995) Protein-structure prediction by threading
  methods - evaluation of current techniques. Proteins 23:337--355.

\bibitem[{Levitt and Warshel(1975)}]{Levitt75_Nature}
Levitt M, Warshel A (1975) Computer simulation of protein folding. Nature
  253:694--8.

\bibitem[{Li {\em et~al.\/}(1996)Li, Helling, Tang, and
  Wingreen}]{Li96_Science}
Li H, Helling R, Tang C, Wingreen N (1996) Emergence of preferred structures in
  a simple model of protein folding. Science 273:666--669.

\bibitem[{Li {\em et~al.\/}(1997)Li, Tang, and Wingreen}]{Li97_PRL}
Li H, Tang C, Wingreen NS (1997) Nature of driving force for protein folding:
  {A} result from analyzing the statistical potential. Phys Rev Lett
  79:765--768.

\bibitem[{Li {\em et~al.\/}(2003)Li, Hu, and Liang}]{Li&Liang03_Proteins}
Li X, Hu C, Liang J (2003) Simplicial edge representation of protein structures
  and alpha contact potential with confidence measure. Proteins 53:792--805.

\bibitem[{Li and Liang(2005{\natexlab{a}})}]{Li&Liang05_PSB}
Li X, Liang J (2005{\natexlab{a}}) Computational design of combinatorial
  peptide library for modulating protein-protein interactions. Pacific
  Symposium of Biocomputing .

\bibitem[{Li and Liang(2005{\natexlab{b}})}]{Li&Liang05_Proteins}
Li X, Liang J (2005{\natexlab{b}}) Geometric cooperativity and
  anti-cooperativity of three-body interactions in native proteins. Proteins
  60:46--65.

\bibitem[{Liang and Dill(2001)}]{LiangDill01_BJ}
Liang J, Dill KA (2001) Are proteins well-packed? Biophys J 81:751--766.

\bibitem[{Liu {\em et~al.\/}(2004)Liu, Zhang, Zhou, and Zhou}]{Zhou04_Proteins}
Liu S, Zhang C, Zhou H, Zhou Y (2004) A physical reference state unifies the
  structure-derived potential of mean force for protein folding and binding.
  Proteins 56:93--101.

\bibitem[{Looger {\em et~al.\/}(2003)Looger, Dwyer, Smith, and
  Hellinga}]{Looger03_Nature}
Looger LL, Dwyer MA, Smith JJ, Hellinga HW (2003) Computational design of
  receptor and sensor proteins with novel functions. Nature 423:185--190.

\bibitem[{Lu and Skolnick(2001)}]{Lu&Skolnick01_Proteins}
Lu H, Skolnick J (2001) A distance-dependent atomic knowledge-based potential
  for improved protein structure selection. Proteins 44:223--232.

\bibitem[{Maiorov and Crippen(1992)}]{Maiorov&Crippen92_JMB}
Maiorov VN, Crippen GM (1992) Contact potential that pecognizes the correct
  folding of globular proteins. J Mol Biol 227:876--888.

\bibitem[{McConkey {\em et~al.\/}(2003)McConkey, Sobolev, and
  Edelman}]{McConkey03_PNAS}
McConkey BJ, Sobolev V, Edelman M (2003) {Discrimination of native protein
  structures using atom-atom contact scoring}. Proc Natl Acad Sci
  100:3215--3220.

\bibitem[{M\'{e}ndez {\em et~al.\/}(2005)M\'{e}ndez, Leplae, Lensink, and
  Wodak}]{CAPRI}
M\'{e}ndez R, Leplae R, Lensink MF, Wodak SJ (2005) Assessment of capri
  predictions in rounds 3-5 shows progress in docking procedures. Proteins
  60:150--169.

\bibitem[{M\'esz\'aros(1996)}]{Meszaros96_CMA}
M\'esz\'aros CS (1996) Fast {C}holesky factorization for interior point methods
  of linear programming. Comp Math Appl 31:49 -- 51.

\bibitem[{Micheletti {\em et~al.\/}(2001)Micheletti, Seno, Banavar, and
  Maritan}]{Micheletti01_Proteins}
Micheletti C, Seno F, Banavar JR, Maritan A (2001) Learning effective amino
  acid interactions through iterative stochastic techniques. Proteins
  42(3):422--431.

\bibitem[{Mirny and Shakhnovich(1996)}]{Mirny&Shakhnovich96_JMB}
Mirny LA, Shakhnovich EI (1996) How to derive a protein folding potential? a
  new approach to an old problem. J Mol Biol 264:1164--1179.

\bibitem[{Mitchell {\em et~al.\/}(1999)Mitchell, Laskowski, Alex, and
  Thornton}]{Thornton99_JCC}
Mitchell BO, Laskowski RA, Alex A, Thornton JM (1999) {BLEEP: potential of mean
  force describing protein-ligand interactions: II. Calculation of binding
  energies and comparison with experimental data}. J Comp Chem 20:1177--85.

\bibitem[{Miyazawa and Jernigan(1985)}]{Miyazawa&Jernigan85_M}
Miyazawa S, Jernigan RL (1985) Estimation of effective interresidue contact
  energies from protein crystal structures: quasi-chemical approximation.
  Macromolecules 18:534--552.

\bibitem[{Miyazawa and Jernigan(1996)}]{Miyazawa&Jernigan96_JMB}
Miyazawa S, Jernigan RL (1996) Residue-residue potentials with a favorable
  contact pair term and an unfavorable high packing density term. J Mol Biol
  256:623--644.

\bibitem[{Miyazawa and Jernigan(2005)}]{Miyazawa&Jernigan05_JCP}
Miyazawa S, Jernigan RL (2005) How effective for fold recognition is a
  potential of mean force that includes relative orientations between
  contacting residues in proteins? The Journal of Chemical Physics 122:024901.

\bibitem[{Momany {\em et~al.\/}(1975)Momany, McGuire, Burgess, and
  Scheraga}]{Scheraga75_JPC}
Momany FA, McGuire RF, Burgess AW, Scheraga HA (1975) {Energy parameters in
  polypeptides. VII. Geometric parameters, partial atomic charges, nonbonded
  interactions, hydrogen bond interactions, and intrinsic torsional potentials
  for the naturally occurring amino acids}. J Phys Chem 79(22):2361--2381.

\bibitem[{Muegge and Martin(1999)}]{Muegge&Martin99_JMC}
Muegge I, Martin YC (1999) A general and fast scoring function for
  protein-ligand interactions: a simplified potential approach. J Med Chem
  42:791--804.

\bibitem[{Munson and Singh(1997)}]{Munson97}
Munson PJ, Singh RK (1997) Statistical significane of hierarchical multi-body
  potential based on delaunay tessellation and their application in
  sequence-structure alignment. Protein Sci 6:1467--1481.

\bibitem[{Nishikawa and Matsuo(1993)}]{Nishikawa&Matsuo93_ProEng}
Nishikawa K, Matsuo Y (1993) {Development of pseudoenergy potentials for
  assessing protein 3-D-1-D compatibility and detecting weak homologies}.
  Protein Eng 6:811--820.

\bibitem[{Papadimitriou and Steiglitz(1998)}]{Papadimitriou98}
Papadimitriou C, Steiglitz K (1998) Combinatorial optimization: algorithms and
  complexity. Dover.

\bibitem[{Park and Levitt(1996)}]{ParkLevitt96_JMB}
Park BH, Levitt M (1996) Energy functions that discriminate x-ray and
  near-native folds from well-constructed decoys. J Mol Biol 258:367--392.

\bibitem[{Park {\em et~al.\/}(2004)Park, Elsner, Staritzbichler, and
  Helms}]{Park_Proteins04}
Park Y, Elsner M, Staritzbichler R, Helms V (2004) Novel scoring function for
  modeling structures of oligomers of transmembrane alpha-helices. Proteins
  57(3):577--85.

\bibitem[{Rank and Baker(1997)}]{Rank&Baker97_ProSci}
Rank JA, Baker D (1997) A desolvation barrier to hydrophobic cluster formation
  may contribute to the rate-limiting step in protein folding. Protein Sci
  6(2):347--354.

\bibitem[{Rohl {\em et~al.\/}(2004)Rohl, Strauss, Misura, and
  Baker}]{Rohl&Baker04_ME}
Rohl CA, Strauss CE, Misura KM, Baker D (2004) Protein structure prediction
  using rosetta. Methods Enzymol 383:66--93.

\bibitem[{Rossi {\em et~al.\/}(2001)Rossi, Micheletti, Seno, and
  Maritan}]{Rossi02_BJ}
Rossi A, Micheletti C, Seno F, Maritan A (2001) A self-consistent
  knowledge-based approach to protein design. Biophys J 80(1):480--490.

\bibitem[{Russ and Ranganathan(2002)}]{proteinDesignReview}
Russ WP, Ranganathan R (2002) Knowledge-based potential functions in protein
  design. Curr Opin Struct Biol 12:447--452.

\bibitem[{Sale {\em et~al.\/}(2004)Sale, Faulon, Gray, and
  Schoeniger}]{Sale_PS04}
Sale K, Faulon J, Gray G, Schoeniger J.S. an d~Young M (2004) Optimal bundling
  of transmembrane helices using sparse distance constraints. Protein Sci
  13(10):2613--27.

\bibitem[{Samudrala and Levitt(2000)}]{DecoyRus}
Samudrala R, Levitt M (2000) {Decoys 'R' Us: a database of incorrect
  conformations to improve protein structure prediction}. Protein Sci
  9:1399--1401.

\bibitem[{Samudrala and Moult(1998)}]{Samudrala&Moult98_JMB}
Samudrala R, Moult J (1998) An all-atom distance-dependent conditional
  probability discriminatory function for protein structure prediction. J Mol
  Biol 275:895--916.

\bibitem[{Sch\"olkopf and Smola(2002)}]{ScholkopfSmola02}
Sch\"olkopf B, Smola AJ (2002) Learning with kernels: {S}upport vector
  machines, regularization, optimization, and beyond. The MIT Press, Cambridge,
  MA.

\bibitem[{{Shakhnovich}(1994)}]{Shakhnovich94_PRL}
{Shakhnovich} EI (1994) Proteins with selected sequences fold into unique
  native conformation. Phys Rev Lett 72:3907--3910.

\bibitem[{Shakhnovich and Gutin(1993)}]{ShakhGutin93_PNAS}
Shakhnovich EI, Gutin AM (1993) Engineering of stable and fast-folding
  sequences of model proteins. Proc Natl Acad Sci USA 90:7195--7199.

\bibitem[{Shimizu and Chan(2001)}]{Shimizu&Chan01_JCP}
Shimizu S, Chan HS (2001) Anti-cooperativity in hydrophobic interactions: A
  simulation study of spatial dependence of three-body effects and beyond. J
  Chem Phys 115(3):1414--1421.

\bibitem[{Shimizu and Chan(2002)}]{Shimizu&Chan02_Proteins}
Shimizu S, Chan HS (2002) Anti-cooperativity and cooperativity in hydrophobic
  interactions: Three-body free energy landscapes and comparison with
  implicit-solvent potential functions for proteins. Proteins 48:15--30.

\bibitem[{Simons {\em et~al.\/}(1999{\natexlab{a}})Simons, Ruczinski,
  Kooperberg, Fox, Bystroff, and Baker}]{Simons99_Proteins}
Simons KT, Ruczinski I, Kooperberg C, Fox B, Bystroff C, Baker D
  (1999{\natexlab{a}}) Improved recognition of native-like protein structures
  using a combination of sequence-dependent and sequence-independent features
  of proteins. Proteins 34:82--95.

\bibitem[{Simons {\em et~al.\/}(1999{\natexlab{b}})Simons, Ruczinski,
  Kooperberg, Fox, Bystroff, and Baker}]{Simons&Baker99_Proteins}
Simons KT, Ruczinski I, Kooperberg C, Fox B, Bystroff C, Baker D
  (1999{\natexlab{b}}) Improved recognition of native-like protein structures
  using a combination of sequence-dependent and sequence-independent features
  of proteins. Proteins 34:82--95.

\bibitem[{Singh {\em et~al.\/}(1996{\natexlab{a}})Singh, Tropsha, and
  Vaisman}]{Singh&Tropsha96_JCB}
Singh RK, Tropsha A, Vaisman II (1996{\natexlab{a}}) Delaunay tessellation of
  proteins: four body nearest-neighbor propensities of amino acid residues. J
  Comput Biol 3(2):213--221.

\bibitem[{Singh {\em et~al.\/}(1996{\natexlab{b}})Singh, Tropsha, and
  Vaisman}]{Singh96_JCB}
Singh RK, Tropsha A, Vaisman II (1996{\natexlab{b}}) Delaunay tessellation of
  proteins: four body nearest-neighbor propensities of amino-acid residues. J
  Comp Bio 3:213--221.

\bibitem[{Sippl(1990)}]{Sippl90_JMB}
Sippl MJ (1990) calculation of conformational ensembles from potentials of the
  main force. J Mol Biol 213:167--180.

\bibitem[{Sippl(1993)}]{Sippl93_JCAMD}
Sippl MJ (1993) Boltzmann's principle, knowledge-based mean fields and protein
  folding. an approach to the computational determination of protein
  structures. J Comput Aided Mol Des 7(4):473--501.

\bibitem[{Sippl(1995)}]{Sippl95_COSB}
Sippl MJ (1995) Knowledge-based potentials for proteins. Curr Opin Struct Biol
  5(2):229--235.

\bibitem[{Tanaka and Scheraga(1976{\natexlab{a}})}]{TanakarScheraga76}
Tanaka S, Scheraga HA (1976{\natexlab{a}}) Medium- and long-range interaction
  parameters between amino acids for predicting three-dimensional structures of
  proteins. Macromolecules 9:945--950.

\bibitem[{Tanaka and Scheraga(1976{\natexlab{b}})}]{Scheraga76_M}
Tanaka S, Scheraga HA (1976{\natexlab{b}}) Medium- and long-range interaction
  parameters between amino acids for predicting three-dimensional structures of
  proteins. Macromolecules 9:945--950.

\bibitem[{Thomas and Dill(1996{\natexlab{a}})}]{Thomas&Dill96_PNAS}
Thomas PD, Dill KA (1996{\natexlab{a}}) {An iterative method for extracting
  energy-like quantities from protein structures}. PNAS 93:11628--33.

\bibitem[{Thomas and Dill(1996{\natexlab{b}})}]{Thomas&Dill96_JMB}
Thomas PD, Dill KA (1996{\natexlab{b}}) Statistical potentials extracted from
  protein structures: {H}ow accurate are they? J Mol Biol 257:457--469.

\bibitem[{Tobi {\em et~al.\/}(2000{\natexlab{a}})Tobi, Shafran, Linial, and
  Elber}]{Tobi&Elber00_Proteins_1}
Tobi D, Shafran G, Linial N, Elber R (2000{\natexlab{a}}) On the design and
  analysis of protein folding potentials. Proteins 40:71--85.

\bibitem[{Tobi {\em et~al.\/}(2000{\natexlab{b}})Tobi, Shafran, Linial, and
  Elber}]{Tobi&Elber00_Proteins}
Tobi D, Shafran G, Linial N, Elber R (2000{\natexlab{b}}) On the design and
  analysis of protein folding potentials. Proteins 40:71--85.

\bibitem[{Vanderbei(1996)}]{Vanderbei96}
Vanderbei RJ (1996) Linear Programming: Foundations and Extensions. Kluwer
  Academic Publishers.

\bibitem[{Vapnik(1995)}]{Vapnik95}
Vapnik V (1995) The Nature of Statistical Learning Theory. Springer, N.Y., ISBN
  0-387-94559-8.

\bibitem[{Vapnik and Chervonenkis(1964)}]{VapChe64}
Vapnik V, Chervonenkis A (1964) A note on one class of perceptrons. Automation
  and Remote Control 25.

\bibitem[{Vapnik and Chervonenkis(1974)}]{VapChe74}
Vapnik V, Chervonenkis A (1974) Theory of Pattern Recognition [in Russian].
  Nauka, Moscow, (German Translation: W.~Wapnik \& A.~Tscherwonenkis, {\em
  Theorie der Zeichenerkennung}, Akademie--Verlag, Berlin, 1979).

\bibitem[{Venclovas {\em et~al.\/}(2003)Venclovas, Zemla, Fidelis, and
  Moult}]{CASP}
Venclovas E, Zemla A, Fidelis K, Moult J (2003) {Comparison of performance in
  successive CASP experiments}. Proteins 45:163--170.

\bibitem[{Vendruscolo and Domanyi(1998)}]{Vendruscolo98_JCP}
Vendruscolo M, Domanyi E (1998) Pairwise contact potentials are unsuitable for
  protein folding. J Chem Phys 109:11101--8.

\bibitem[{Vendruscolo {\em et~al.\/}(2000{\natexlab{a}})Vendruscolo,
  Najmanovich, and Domany}]{Vendruscolo00_Proteins}
Vendruscolo M, Najmanovich R, Domany E (2000{\natexlab{a}}) Can a pairwise
  contact potential stabilize native protein folds against decoys obtained by
  threading? Proteins Structure Function and Genetics 38:134--148.

\bibitem[{Vendruscolo {\em et~al.\/}(2000{\natexlab{b}})Vendruscolo,
  Najmanovich, and Domany}]{Vendruscolo00_Proteins.pair}
Vendruscolo M, Najmanovich R, Domany E (2000{\natexlab{b}}) Can a pairwise
  contact potential stabilize native protein folds against decoys obtained by
  threading? Proteins Structure Function and Genetics 38:134--148.

\bibitem[{Wodak and Rooman(1993)}]{Wodak93_COSB}
Wodak SJ, Rooman MJ (1993) Generating and testing protein folds. Curr Opin
  Struct Biol 3:247--259.

\bibitem[{Wolynes {\em et~al.\/}(1995)Wolynes, Onuchic, and
  Thirumalai}]{Wolynes95_Science}
Wolynes PG, Onuchic JN, Thirumalai D (1995) Navigating the folding routes.
  Science 267:1619--20.

\bibitem[{Xia and Levitt(2000)}]{Xia&Levitt00_JCP}
Xia Y, Levitt M (2000) Extracting knowledge-based energy functions from protein
  structures by error rate minimization: Comparison of methods using lattice
  model. J Chem Phys 113:9318--9330.

\bibitem[{Xu {\em et~al.\/}(1997)Xu, Lin, and Nussinov}]{Xu&Nussinov97_JMB}
Xu D, Lin SL, Nussinov R (1997) Protein binding versus protein folding: the
  role of hydrophilic bridges in protein associations. J Mol Biol
  265{1}:68--84.

\bibitem[{Zhang and Kim(2000)}]{Zhang&Kim00_PNAS}
Zhang C, Kim SH (2000) {Environment-dependent residue contact energies for
  proteins}. PNAS 97:2550--2555.

\bibitem[{Zhang {\em et~al.\/}(2004{\natexlab{a}})Zhang, Liu, Zhou, and
  Zhou}]{Zhang&Zhou04_ProSci}
Zhang C, Liu S, Zhou H, Zhou Y (2004{\natexlab{a}}) {An accurate,
  residue-level, pair potential of mean force for folding and binding based on
  the distance-scaled, ideal-gas reference state}. Protein Sci 13:400--411.

\bibitem[{Zhang {\em et~al.\/}(2004{\natexlab{b}})Zhang, Liu, Zhou, and
  Zhou}]{Zhang_BJ04}
Zhang C, Liu S, Zhou H, Zhou Y (2004{\natexlab{b}}) The dependence of all-atom
  statistical potentials on structural training database. Biophys J
  86(6):3349--58.

\bibitem[{Zhang {\em et~al.\/}(2004{\natexlab{c}})Zhang, Liu, and
  Zhou}]{Zhou04_BiophysJ}
Zhang C, Liu S, Zhou Y (2004{\natexlab{c}}) The dependence of all-atom
  statistical potentials on training structural database. Biophys J
  86:3349--3358.

\bibitem[{Zhang {\em et~al.\/}(2005)Zhang, Liu, Zhu, and
  Zhou}]{Zhang&Zhou05_JMC}
Zhang C, Liu S, Zhu Q, Zhou Y (2005) A knowledge-based energy function for
  protein-ligand, protein-protein, and protein-dna complexes. J Med Chem
  48:2325--35.

\bibitem[{Zhang {\em et~al.\/}(1997)Zhang, Vasmatzis1, Cornette, and
  DeLisi}]{Zhang&Delisi97_JMB}
Zhang C, Vasmatzis1 G, Cornette JL, DeLisi C (1997) Determination of atomic
  desolvation energies from the structures of crystallized proteins. J Mol Biol
  267.

\bibitem[{Zheng {\em et~al.\/}(1997{\natexlab{a}})Zheng, Cho, Vaisman, and
  Tropsha}]{Zheng&Tropsha97_PSB}
Zheng W, Cho SJ, Vaisman II, Tropsha A (1997{\natexlab{a}}) A new approach to
  protein fold recognition based on delaunay tessellation of protein structure.
  Pac Symp Biocomput :486--497.

\bibitem[{Zheng {\em et~al.\/}(1997{\natexlab{b}})Zheng, Cho, Vaisman, and
  Tropsha}]{Zheng97}
Zheng W, Cho SJ, Vaisman II, Tropsha A (1997{\natexlab{b}}) A new approach to
  protein fold recognition based on {D}elaunay tessellation of protein
  structure. In: Altman R, Dunker A, Hunter L, Klein T (eds.), Pacific
  Symposium on Biocomputing'97, World Scientific, Singapore, pp. 486--497.

\bibitem[{Zhou and Zhou(2002)}]{Zhou02_ProSci}
Zhou H, Zhou Y (2002) Distance-scaled, finite ideal-gas reference state
  improves structure-derived potentials of mean force for structure selection
  and stability prediction. Protein Sci 11:2714--2726.

\end{thebibliography}
\end{document}